\documentclass[letterpaper]{article} % DO NOT CHANGE THIS
\usepackage{aaai2027}

\usepackage[hyphens]{url}  % DO NOT CHANGE THIS
\usepackage{graphicx} % DO NOT CHANGE THIS
\usepackage{natbib}  % DO NOT CHANGE THIS AND DO NOT ADD ANY OPTIONS TO IT
\usepackage{caption} % DO NOT CHANGE THIS AND DO NOT ADD ANY OPTIONS TO IT
\usepackage{algorithm}
\usepackage{algorithmic}

\usepackage{newfloat}
\usepackage{listings}
\DeclareCaptionStyle{ruled}{labelfont=normalfont,labelsep=colon,strut=off} % DO NOT CHANGE THIS
\floatstyle{ruled}
\newfloat{listing}{tb}{lst}{}
\floatname{listing}{Listing}

\usepackage{booktabs}
\usepackage{amsmath}
\usepackage{amssymb}

\usepackage{booktabs}
\usepackage{array}
\usepackage{xcolor}
\usepackage{colortbl}
\usepackage{multirow}
\usepackage[table]{xcolor}

\usepackage{longtable}

\definecolor{apibanner}{HTML}{70FFFF}    % API分组蓝绿背景
\definecolor{openbanner}{HTML}{99FF99}   % 开源分组嫩绿背景
\definecolor{bestgreen}{HTML}{D9F2D9}    % 优秀数值绿高亮
\definecolor{worstred}{HTML}{FFCCCC}     % 差值红高亮

\definecolor{apibanner}{HTML}{70FFFF}    % API分组蓝绿背景
\definecolor{openbanner}{HTML}{99FF99}   % 开源分组嫩绿背景
\definecolor{bestgreen}{HTML}{D9F2D9}    % 优秀数值绿高亮
\definecolor{worstred}{HTML}{FFCCCC}     % 差值红高亮
\definecolor{AppBlue}{HTML}{4B86B4}
\definecolor{AppFill}{HTML}{F4F7FA}

\definecolor{AppBand}{HTML}{E8E8E8}
\definecolor{AppStripe}{HTML}{F6F6F6}
\definecolor{AppLine}{HTML}{707070}



\newcommand{\ShowcasePanel}[2]{%

  \begingroup

  \setlength{\fboxrule}{0.9pt}%

  \setlength{\fboxsep}{0pt}%

  \noindent\fcolorbox{AppBlue}{AppFill}{%

    \begin{minipage}{\dimexpr\linewidth-2\fboxrule\relax}

      \begingroup

      \setlength{\fboxsep}{5pt}%

      \colorbox{AppBlue}{%

        \parbox{\dimexpr\linewidth-2\fboxsep\relax}{%

          \color{white}\sffamily\bfseries #1}}%

      \endgroup

      \par\smallskip

      \noindent\hspace{6pt}%

      \begin{minipage}{\dimexpr\linewidth-12pt\relax}

        \footnotesize #2

      \end{minipage}

      \smallskip

    \end{minipage}}%

  \endgroup

  \par

} 
\title{SocialBuddy: Tailoring Search Agent for Social Media Ecosystems}
\author{
    Mingxuan Li\textsuperscript{\rm 1,2} \equalcontrib,
    Yirong Mao\textsuperscript{\rm 1} \equalcontrib,
    FaZhan Zhang\textsuperscript{\rm 2},
    Haibiao Yao\textsuperscript{\rm 3},
    Runze Hu\textsuperscript{\rm 2},
    Wenhui Que\textsuperscript{\rm 1}\corresponding
}
\affiliations{
    \textsuperscript{\rm 1}WeChat, Tencent Inc, China\\
    \textsuperscript{\rm 2}School of Information and Electronics, Beijing Institute of Technology, China\\
    \textsuperscript{\rm 3}School of Computer Science and Technology, University of Science and Technology of China, China\\
    limx1630@gamil.com,
    \{minsenli,erongmao,victorque\}@tencent.com
  
}

\begin{document}

\maketitle

\begin{abstract}
% 等结果出来后面把最后的一句改改
In the era of digital social interaction, searching friends' posts from massive social streams has become a fundamental user need. However, while modern agentic search frameworks have achieved remarkable success in conventional retrieval tasks, they break down when confronted with heterogeneous user queries and multi-dimensional social feeds, resulting in severe performance degradation in complex social search. To bridge this gap, we introduce \textbf{SocialBuddy}, the first agentic search framework tailored for social scenarios. Specifically, we construct \textbf{SocialEnv}, the first large-scale simulated environment for social search. Powered by an automated data and trajectory synthesis pipeline, SocialEnv includes 200K user profiles, 10 million social posts, and 50K reasoning trajectories, establishing a solid foundation for the development of social search agents. To tackle the credit assignment dilemma caused by sparse rewards in social search, we design \textbf{SocialPO}, a hybrid-granularity optimization framework. It macroscopically reinforces successful reasoning paths via multi-dimensional rewards, while microscopically rectifying deviated trajectories through fine-grained prefix truncation and token-level supervision. This hybrid-granularity design delivers multi-scale guidance in complex long-sequence scenarios. Finally, we construct \textbf{SocialSearch Benchmark} to provide a quantitative evaluation scheme for assessing the social search capabilities of SocialBuddy. Extensive experiments demonstrate that SocialBuddy-35B surpasses significantly larger frontier LLMs. Code and dataset will be released upon article acceptance.

\end{abstract}


\section{Introduction}
Large Language Models (LLMs) have emerged as a cornerstone of modern artificial intelligence, demonstrating unprecedented capabilities in complex reasoning \cite{brown2020language,wei2022cot}. However, their parametric knowledge remains bounded by static training data and strict cutoff dates. As a result, when confronted with dynamic, real-time, or proprietary contexts, LLMs frequently struggle with knowledge staleness and exhibit a high risk of hallucination \cite{lewis2020rag,ji2023survey,chen2024rgb,song2025smartsearcher}.

\begin{figure}[t]
\begin{center}
 \includegraphics[width=\columnwidth]{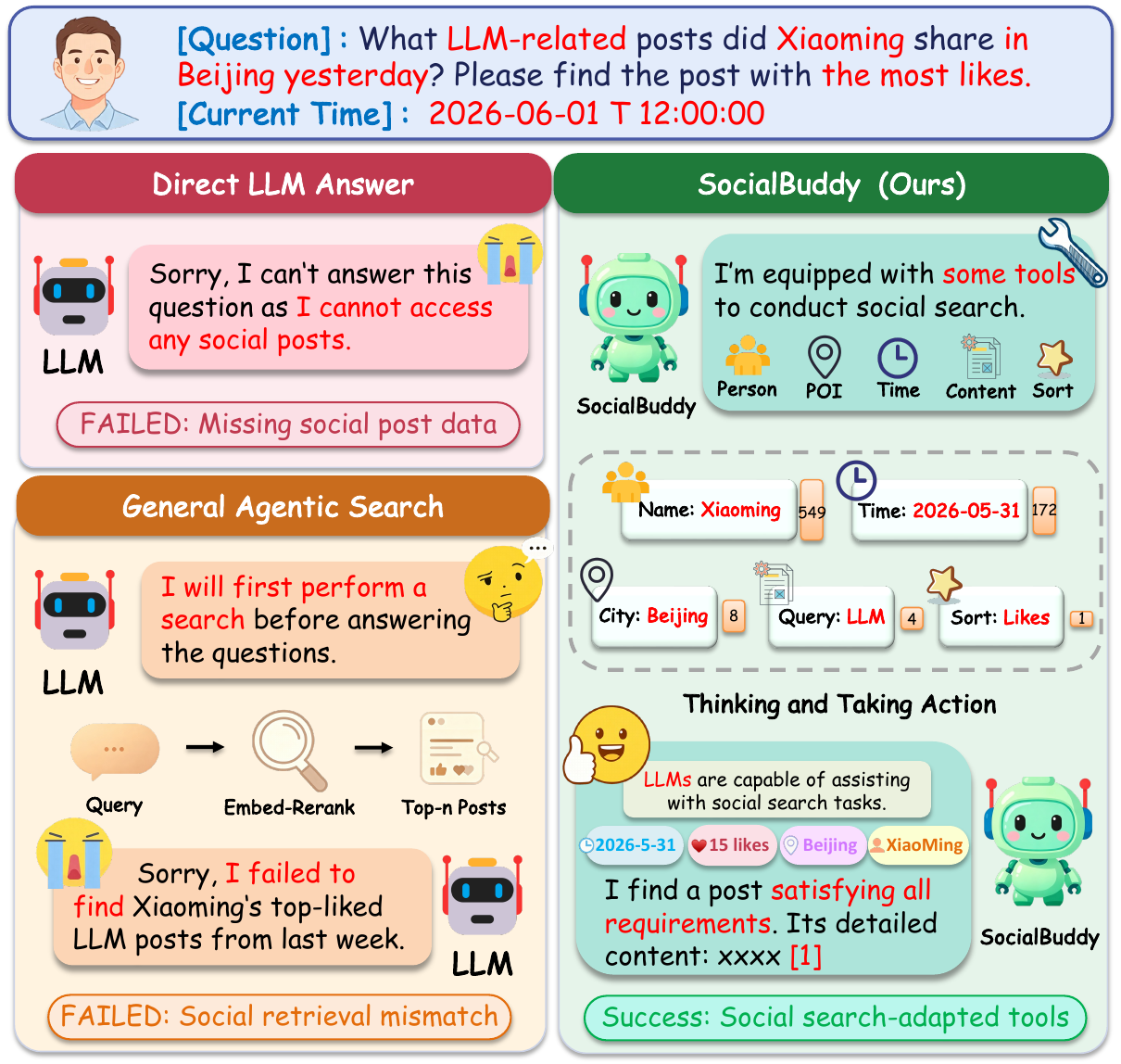}
\end{center}
\caption{A comparison of social search approaches. \textbf{Left:} Standard LLMs and general search agents fail due to data access barriers and the mismatch between text retrieval and multi-dimensional social constraints. \textbf{Right:} SocialBuddy employs social-adapted tools to decompose complex queries and perform relational reasoning over structured metadata.}
\label{fig1}
\end{figure} 

% By empowering models to autonomously plan, execute multi-turn queries, and adaptively integrate external knowledge, these studies have achieved state-of-the-art performance in complex retrieval tasks \cite{li2025searcho1,zheng2025deepresearcher,li2026websailorv2}. 
To overcome these inherent limitations, a surge of recent research has proposed various agentic search frameworks \cite{nakano2021webgpt,schick2023toolformer, chen2025mindsearch}. These frameworks leverage LLMs as central decision-makers to autonomously orchestrate external tools like search engines, databases, and retrievers, thereby dynamically fetching up-to-date information to augment the reasoning process. Prominent instantiations include text-based (e.g., Search-R1 \cite{jin2025searchr1}) and multimodal agentic search frameworks (e.g., MMSearch-R1 \cite{wu2026mmsearchr1}).

% Prominent instantiations include text-based frameworks such as
% Search-R1~\cite{jin2025searchr1}, and multimodal search frameworks
% such as SearchLVLMs,
% ReFAct~\cite{wu2026refact}, and
% MMSearch-R1~\cite{wu2026mmsearchr1}.

Despite their remarkable success in conventional open-domain web searches \cite{li2024searchlvlms,wu2026refact}, existing frameworks have surprisingly overlooked the most ubiquitous and content-rich application scenario in daily human life: social media ecosystems. Every day, billions of users navigating platforms like WeChat Moments, Facebook, and X (formerly Twitter) share a vital yet unaddressed need \cite{tencent2026q1,meta2026q1,morris2010social}. They need to efficiently search, track, and synthesize their friends' multi-dimensional posts to answer the intuitive question, "What are my friends up to?" However, as illustrated in Fig.\ref{fig1} (left), current agentic search frameworks are fundamentally inapplicable to these unique environments.

%%%%这里加一下首页图
This failure stems from two fundamental misalignments between conventional agentic search and the unique requirements of social media scenarios. First, conventional embedding-based retrieval mechanisms are fundamentally misaligned with the implicit and relational nature of social queries \cite{shu2026soulseek}. 
Consider a representative query: "What LLM-related posts did Xiaoming share on Moments in the past week? Find the one with the most likes." Traditional search frameworks often fail to isolate "Xiaoming" as a specific user or dynamically filter posts by the "past week" time window, leading to severe semantic drift and flooding the context window with irrelevant data~\cite{hsia2025ragged}. Moreover, unlike static search indexes, social feeds exhibit a multi-dimensional architecture combining content, user profiles, timestamps, and dynamic engagement metrics~\cite{xue2026some,su2025facebookscoped}. Determining constraints like "the one with the most likes" requires joint quantitative reasoning over these relational metadata, which far exceeds the capacity of standard retrievers. Without specialized multi-attribute filtering tools, general-purpose agents struggle to navigate such interleaved social environments.

To fill this critical void, \textbf{we present SocialBuddy, the first agentic search framework tailored specifically for social media environments}, as illustrated in Fig.\ref{fig1} (right). Distinct from conventional agentic search, SocialBuddy redefines agentic search in social domains by coupling semantic understanding with multi-dimensional context parsing. To achieve this, we equip SocialBuddy with a specialized suite of tools, allowing the agent to seamlessly search, filter, and aggregate posts across key social dimensions. 

In order to develop Socialbuddy, \textbf{we construct SocialEnv, the first large-scale simulated environment for social agentic search.} SocialEnv realistically models social networks, encompassing 200K users, 10 million posts, and 50K interaction trajectories, thereby providing a robust environmental and data foundation for agent training and multi-turn reasoning optimization. Crucially, navigating such complex, long-horizon search trajectories under sparse environmental feedback is highly prone to severe credit assignment failures during policy learning. To resolve this, \textbf{we present SocialPO, a hybrid-granularity policy optimization framework.} SocialPO macroscopically reinforces successful reasoning paths via joint multi-dimensional outcome rewards, while microscopically rectifying deviated trajectories through fine-grained prefix truncation and token-level supervision. This hybrid-granularity design bridges holistic outcome alignment and localized step-level corrections, delivering precise, multi-scale guidance in complex long-sequence scenarios. Moreover, \textbf{we present the SocialSearch Benchmark}, providing an accompanying, systematic evaluation protocol to quantitatively assess the social search capabilities of SocialBuddy.

In summary, our main contributions are as follows:
\begin{itemize}
\item \textbf{Pioneering Social Agentic Search Framework:} We propose \textbf{SocialBuddy}, the first agentic search framework tailored specifically for social media ecosystems, effectively extending the capabilities of conventional agentic search into the realm of daily social intelligence.
\item 
\textbf{Environment \& Algorithm:} We construct the first large-scale simulated social environment \textbf{SocialEnv} and its comprehensive data pipeline, providing a high-fidelity simulation environment and robust training data for social search. To tackle the severe credit assignment dilemma under sparse, long-horizon rewards, we introduce \textbf{SocialPO}, a hybrid-granularity policy optimization framework designed to simultaneously reinforce valid decisions and rectify localized errors.
\item
\textbf{Benchmark \& Evaluation:} To systematically evaluate the social search and summarization capabilities of SocialBuddy, we establish the \textbf{SocialSearch Benchmark}. Extensive experiments demonstrate that our proposed SocialBuddy-35B consistently outperforms significantly larger, top-tier LLM baselines.

\end{itemize}

% Large Language Models (LLMs) have fundamentally redefined the landscape of natural language processing, demonstrating unprecedented capabilities across a broad spectrum of understanding and generation tasks. However, the parametric knowledge embedded within these models is inherently bounded by static training data and strict knowledge cutoff dates. As a result, LLMs inevitably suffer from critical limitations such as knowledge obsolescence, factual gaps, and hallucinations, particularly when confronted with dynamic, real-time, or highly proprietary information. To transcend these boundaries, recent advancements have increasingly capitalized on the exceptional tool-utilization capabilities of LLMs. This paradigm shift has fueled the proliferation of various agentic search frameworks. By autonomously calling external tools—such as search engines, database APIs, and retrievers—these agents can dynamically fetch up-to-date and context-rich information, thereby augmenting the model's reasoning process and establishing a more factual foundation for open-domain information retrieval tasks.

\begin{figure*}[t]
\begin{center}
 \includegraphics[width=\textwidth]{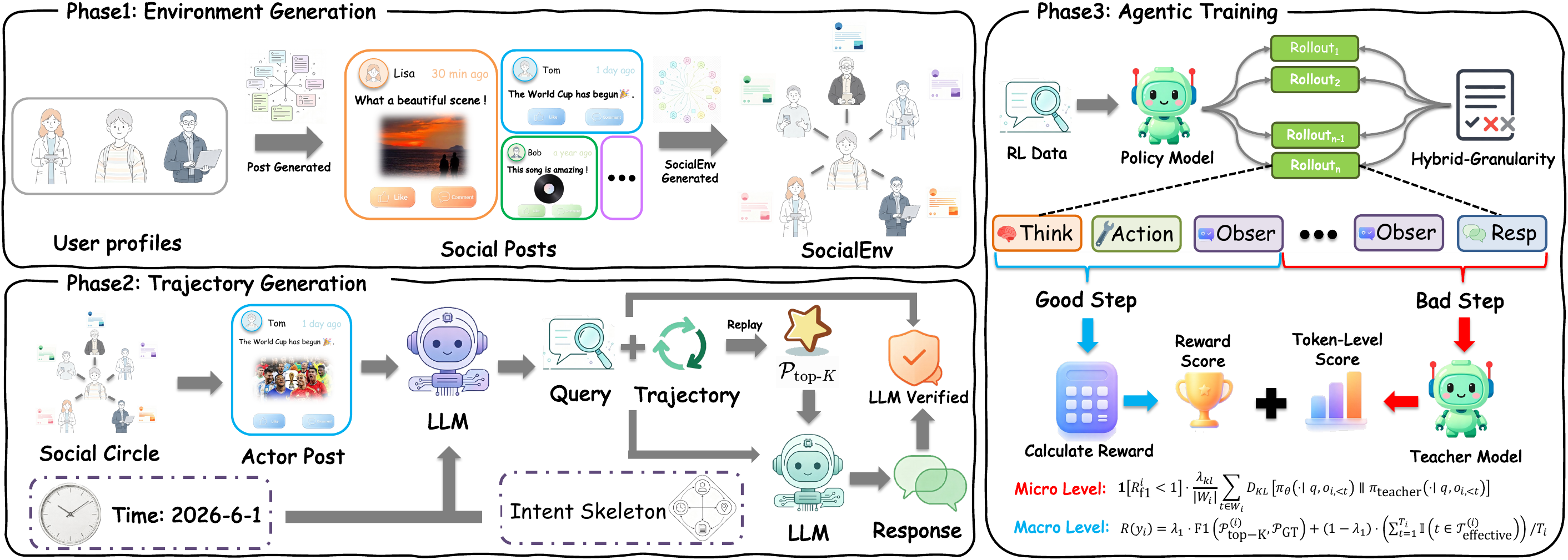}
\end{center}
\caption{The Overall framework of SocialBuddy with three main phases. Phase 1: Building the interactive SocialEnv; Phase 2: Curating multi-intent query-trajectory pairs; and Phase 3: Optimizing SocialBuddy via the SocialPO algorithm.}
\label{fig2}
\end{figure*}

\section{Related Work}   %v2
\subsection{Retrieval-Augmented Generation}
Retrieval-Augmented Generation (RAG) grounds language models in external knowledge bases to mitigate knowledge staleness and hallucinations~\cite{lewis2020rag}. Early systems combine dense or late-interaction retrievers with a static retrieve-then-generate pipeline~\cite{karpukhin2020dense, khattab2020colbert, izacard2021leveraging}. To enhance decision-making flexibility, recent studies introduce adaptive retrieval mechanisms. For instance, FLARE anticipates upcoming information needs during generation, Self-RAG incorporates self-reflection to critique retrieved evidence on demand, and CoRAG interleaves dynamic retrieval with chain-of-thought reasoning~\cite{jiang2023active, asai2023selfrag, wang2026chain}. 

However, existing RAG paradigms target document-level search and struggle with complex multi-intent queries in social environments. Social search requests typically involve heterogeneous constraints, such as user identities, time windows, locations, and content types. Standard RAG mechanisms often collapse these constraints into a single query vector, causing constraint drift and information loss.

%Social queries instead impose conjunctive, user-dependent constraints—for example, matching a topic while restricting the author to a friend, a recent time window, and posts visible to the requester. SocialBuddy addresses this retrieval mismatch by maintaining a user- and time-conditioned post pool and using explicit relational, temporal, spatial, content, and engagement operations rather than relying on one-shot embedding similarity.

\subsection{Agentic Search}
Search agents turn information seeking into an iterative reasoning--action process. WebGPT and ReAct established tool-mediated interaction~\cite{nakano2021webgpt, yao2023react}. Search-o1 and Search-R1 subsequently interleave reasoning with multi-turn retrieval~\cite{li2025searcho1, jin2025searchr1}. More recent systems such as DeepResearcher scale reinforcement learning to long-horizon search on the open web~\cite{zheng2025deepresearcher}. Together, these studies show that planning and learned tool use can substantially improve knowledge-intensive reasoning.

However, existing search agents primarily focus on open-web pages, largely ignoring social media platforms which host billions of active users generating massive streams of heterogeneous, multi-attribute posts with strict temporal and relational constraints. Therefore, developing a dedicated framework for social search is of paramount importance.

%Although hybrid social retrieval and SoMe improve platform search or evaluation~\cite{su2025facebookscoped, xue2026some}, they do not train an interactive agent for this full setting. SocialBuddy closes this agent–environment mismatch through SocialEnv, a social-specific five-tool action space, and SocialPO for learning reliable long-horizon trajectories.

\section{Method}
In this section, we present the overall method for SocialBuddy as illustrated in Fig.\ref{fig2}. Section 3.1 outlines the system architecture and interactive action space. Section 3.2 presents the construction of SocialEnv, alongside its trajectory curation pipeline. Section 3.3 details the two-stage training pipeline combining Supervised Fine-Tuning and SocialPO Reinforcement Learning. Finally, Section 3.4 introduces the SocialSearch Benchmark and its evaluation protocol to assess social search capabilities.

\subsection{SocialBuddy Architecture}
\subsubsection{Interaction Formulation.}

Following the ReAct paradigm \cite{yao2023react}, SocialBuddy operates as an iterative reasoning-and-acting agent tailored for dynamic social media ecosystems. Unlike conventional agentic search where the information repository is globally static, social media retrieval is inherently user-centric and highly time-sensitive.

To capture these characteristics, we define a dynamic social environment as $\mathcal{E}_{(\mathcal{U}, \mathcal{T})}$, which is jointly conditioned on the target user $\mathcal{U}$ encapsulating their distinct social circles and the absolute timestamp $\mathcal{T}$ when the query is initiated. Consequently, given a query $q$ triggered by user $\mathcal{U}$ at time $\mathcal{T}$, SocialBuddy interacts with this personalized, time-varying environment to generate a social search trajectory $\tau$:

\begin{equation}
\tau = \left( q, \, (r_0, a_0, o_0^{(\mathcal{U}, \mathcal{T})}), \, \dots, \, (r_T, a_T, o_T^{(\mathcal{U}, \mathcal{T})}), \, y \right).
\end{equation}

At each turn $t$, the agent policy $\pi_\theta$ generates a reasoning trace $r_t$ over the historical context $h_t$, executes a tool action $a_t \sim \pi_\theta(\cdot \mid h_t, r_t)$, and receives an observation $o_t^{(\mathcal{U}, \mathcal{T})} \sim \mathcal{E}_{(\mathcal{U}, \mathcal{T})}(\cdot \mid a_t)$. Crucially, this environment formulation introduces two core divergences from standard search: 1) Spatial Divergence: Disparate users $\mathcal{U}_1, \mathcal{U}_2$ at the same time $\mathcal{T}$ receive variant observations ($o_t^{(\mathcal{U}_1, \mathcal{T})} \neq o_t^{(\mathcal{U}_2, \mathcal{T})}$) due to segregated social circles. 2) Temporal Divergence: The same user $\mathcal{U}$ at different timestamps $\mathcal{T}_1, \mathcal{T}_2$ encounters distinct observations ($o_t^{(\mathcal{U}, \mathcal{T}_1)} \neq o_t^{(\mathcal{U}, \mathcal{T}_2)}$). %due to real-time feed streaming.
%This process iterates until the agent provides a final answer $y$.

\subsubsection{Action Space Formulation.}

To interact with the personalized, time-varying environment $\mathcal{E}_{(\mathcal{U}, \mathcal{T})}$, SocialBuddy defines a structured action space $\mathcal{A}$ comprising five atomic tools that sequentially filter, refine, and rank an active post pool $\mathcal{P}$. Formally, the post pool $\mathcal{P}$ represents the set of all posts visible to the current user $u \in \mathcal{U}$ at the specific timestamp $t \in \mathcal{T}$. Starting from this initial state $P_0 = \mathcal{P}$, each filtering tool $a \in \mathcal{A}$ iteratively drives a state transition from $P_k$ to $P_{k+1}$:

\begin{equation}
P_{k+1} = \{p \in P_k \mid \mathcal{F}_{a}(p,c_{a}) = 1\}.
\end{equation}

where $P_k \subseteq \mathcal{P}$ denotes the active pool at step $k$, and $\mathcal{F}_{a}$ serves as the tool-specific indicator function evaluating whether a post satisfies the constraint parameters $c_{a}$.

%Specifically, the action space $A$ comprises five functional tools, each tailored to resolve a specific dimension of user queries. To handle requests regarding social relationships, $Tool_{user}$ extracts relational or demographic targets into $c_{user}$ to isolate posts based on author profiles, directly resolving queries such as "posts from close colleagues" or "middle-aged male users". Temporal queries are managed by $Tool_{time}$, which maps time-sensitive expressions into $c_{time}$ to target posts from specific chronological windows, such as "from a week ago" or "at midnight during the Dragon Boat Festival". For geographic requests, $Tool_{loc}$ converts spatial descriptions into $c_{loc}$ to pinpoint feeds tied to specific locations, satisfying queries like "currently in Chongqing". Meanwhile, $Tool_{con}$ resolves explicit format and abstract thematic requests by consolidating content criteria into $c_{con}$, allowing SocialBuddy to simultaneously filter structural media types (e.g., "text-only feeds" or "shared articles") and target specific topics (e.g., "AI or electronic music"). Finally, $Tool_{fin}$ resolves user requests regarding presentation ordering and item layout. It ranks the surviving posts by candidate-normalized importance, likes, comments, or total interactions and returns a fixed maximum of five posts. The rationale for these tool boundaries and their interaction protocol is provided in \textcolor{red}{Appendix}.

Specifically, the action space $A$ comprises five tools, each tailored to resolve a specific dimension of user queries. $Tool_{user}$ isolates posts by author demographics or relationships (e.g., "close colleagues"), while $Tool_{time}$ and $Tool_{loc}$ filter chronological windows (e.g., "midnight during the Dragon Boat Festival") and spatial locations (e.g., "currently in Beijing"), respectively. $Tool_{con}$ addresses content criteria, simultaneously filtering media formats (e.g., "shared articles") and abstract topics (e.g., "AI or electronic music"). Finally, $Tool_{fin}$ manages presentation layout, ranking surviving posts by interaction metrics or normalized importance to return at most five items. More details about these tools are provided in Appendix.

By sequentially orchestrating these tools, SocialBuddy systematically translates a social query into deterministic and verifiable execution pipelines.

\subsection{Environment and Trajectory Generation}

\subsubsection{Environmental Simulation.}
Due to strict privacy constraints, real-world social data is largely inaccessible, creating an urgent need for synthesized social environments \cite{fiesler2020robots,yang2025oasisopenagentsocial}. Therefore, we introduce SocialEnv. To capture the intrinsic characteristics of real-world social platforms (e.g., WeChat Moments), SocialEnv abstracts the dynamic social environment into localized, ego-centric subgraphs. Specifically, the social sphere is naturally partitioned into numerous overlapping sub-circles \cite{NIPS2012_7a614fd0}. For any designated central user $u \in \mathcal{U}$, the environment restricts $u$'s observable scope $\mathcal{E}_u^{(\mathcal{T})}$ strictly to their immediate first-degree connections and the content published by these direct peers:

\begin{equation}
\mathcal{E}_u^{(\mathcal{T})} = \Big( \{u\} \cup \mathcal{N}(u), \; \mathcal{P}_{\{u\} \cup \mathcal{N}(u)}^{(\le \mathcal{T})} \Big)
\end{equation}

where $\mathcal{N}(u)$ denotes the set of $u$'s direct first-degree friends, and $\mathcal{P}_{\{u\} \cup \mathcal{N}(u)}^{(\le \mathcal{T})}$ represents the collection of social posts generated by user $u$ and their friends up to time $\mathcal{T}$. Consequently, $u$ possesses zero visibility over second-degree entities. This structural formulation rigorously enforces privacy-aware permission boundaries in social search.

To build the SocialEnv, we establish a systematic three-stage pipeline as shown in Fig.\ref{fig2} (Phase 1).
%We first construct a diverse set of user personas encapsulating distinct demographic traits, personal backgrounds, and interest profiles. Conditioned on these personas, we then generate heterogeneous social posts representing everyday status updates and shared media. Finally, we select central target users and configure their corresponding first-degree friend networks, enforcing localized visibility constraints to finalize the executable environment.
Specifically, we first construct a diverse set of user personas encapsulating distinct demographic traits, personal backgrounds, and interest profiles following  \cite{ge2024scaling}. Crucially, we partition these persona attributes into public and private fields to strictly model the information boundaries inherent in social platforms. During the social post synthesis stage, the complete profile, including private attributes such as family details, personal values, and life attitudes, is injected into the generator to ensure that synthesized posts naturally reflect each user's implicit motivations and daily realities. Conversely, during the social search process, a central user's view of other personas is strictly restricted to observable public attributes, including basic demographic markers (e.g., age, gender), surface identities (e.g., remark name), and social proximity metrics (e.g., relationship type).

Based on user personas, we sample post timestamps aligned with human diurnal patterns \cite{Naaman_Zhang_Brody_Lotan_2021}, draw themes from a topic library, and extract relevant contexts along with distinct communication styles to drive post creation. Powered by a Large Language Model, these sampled elements are synthesized into structured post entries across five distinct post types, including plain text, image, video, article share, and music share. Each entry captures rich metadata, such as timestamps and topic labels. Additionally, posts optionally include location tags, as well as simulated like and comment counts.
%, thereby creating a highly realistic and diverse social stream for every user.

%Based on the designated post type, the schema dynamically incorporates visual descriptions and media counts for original media, or authentic titles, artists, and text excerpts retrieved from real-world platforms for shared articles and music. 

Finally, to assemble the complete executable environment, we construct localized social circles centered around randomly selected anchor users. For each designated central user, a subset of other personas is assigned to establish their immediate first-degree network, thereby simulating visibility constraints of real-world social platforms. In total, we synthesize 200K user profiles and 10 million social posts, partitioning them into 1,000 distinct friend circles. Further implementation details are provided in the Appendix.

\subsubsection{Trajectory Generation.}

A key challenge in synthesizing social search trajectories is ensuring sampling efficiency across localized environments. Randomly prompting a teacher model to query arbitrary social circles often yields empty search results due to the lack of environment priors, leading to massive computational waste. In contrast, real-world user queries naturally carry strong prior knowledge of target contexts. 

To bridge this gap, we propose an anchor-guided inverse trajectory synthesis strategy that reverse-engineers user queries and gold trajectories from observable posts (Figure~\ref{fig2}, Phase 2). Specifically, given an ego-centric circle, we first set a temporal cutoff $t_{\text{now}}$ to restrict feed visibility strictly to historical posts. We then sample an intent skeleton specifying a target combination of key search dimensions, such as person, time, location, and content. Rather than sampling parameters independently, we select a ground-truth anchor post from the observable pool and extract its metadata, including author relationship, timestamp, tagged location, and post topic, to populate the sampled skeleton's tool arguments. This anchor-guided parameter mapping guarantees a non-empty target set containing at least the anchor post itself, while naturally preserving multi-dimensional query complexity. Finally, a strong teacher LLM translates these structured constraints and tool specifications into colloquial user queries and step-by-step reasoning trajectories.

During intent skeleton extraction, we deliberately vary constraint complexity to generate trajectories across diverse difficulty levels, ranging from precise multi-attribute queries (e.g., finding the most-liked post about LLMs published by Xiaoming last Wednesday) to broad single-dimension requests (e.g., finding any posts from yesterday). Because broadly constrained queries naturally retrieve a much larger pool of matching items, we perform a deterministic execution replay across all difficulty tiers to collect the complete candidate set $\mathcal{P}_{\text{recalled}}$. However, since end users rarely require exhaustive enumeration, each trajectory concludes with $\text{Tool}_{\text{fin}}$, which performs Top-$K$ selection over $\mathcal{P}_{\text{recalled}}$. Sorting strictly obeys explicit query constraints (e.g., highest like counts) or defaults to a composite score that balances recency and overall engagement. The complete trajectory is abstracted as:

\begin{equation}
\begin{aligned}
\mathcal{O} &= \left\langle R \sim P_{\theta}(R \mid q, \mathcal{P}_{\text{top-}K}), \quad \mathcal{P}_{\text{top-}K} \right\rangle.
\end{aligned}
\end{equation}

Here, $R$ represents the generated natural language response, while $\mathcal{P}_{\text{top-}K}$ denotes the structured set of final selected target posts. Finally, a quality verification module audits the complete execution trajectory: verifying whether the recalled post set strictly matches the target items specified by the query, and checking whether the generated text response is fluent, natural, and free of hallucinations.

\subsection{Agentic Training}
\subsubsection{Supervised Fine-Tuning.}
The Supervised Fine-Tuning phase optimizes the base model via next-token prediction \cite{NEURIPS2022_b1efde53}. We sample a subset of low-complexity trajectories consisting of simple queries with only two to three intent constraints. Training on these elementary samples guides the model to quickly learn fundamental tool invocation protocols and structured response generation. 
%, providing a stable foundation for handling intricate multi-step reasoning
\subsubsection{Reinforcement Learning.}
Reinforcement learning for social agentic search encounters two critical limitations when relying solely on outcome-based rewards. First, severe reward sparsity leaves long-horizon search trajectories without intermediate feedback, failing to penalize repetitive fetches and yielding bloated, inefficient execution paths. Second, the credit assignment dilemma hinders fine-grained supervision over intermediate actions. Coarse terminal signals fail to isolate local missteps, leaving the policy unable to targetedly rectify erroneous steps within search paths.

To tackle these challenges, we introduce SocialPO, a hybrid-granularity policy optimization framework as as illustrated in Fig.\ref{fig2} (Phase 3). For each query $q$, we sample a group of $G$ candidate trajectories $\{y_i\}_{i=1}^G$ from the old policy $\pi_{\theta_{\text{old}}}$. At the macro level, to alleviate reward sparsity, the quality of each trajectory $y_i$ is evaluated via a composite reward $R(y_i)$ combining outcome accuracy with the ratio of effective tool calls:

\begin{equation}
R(y_i) = \lambda \cdot  \text{F1}(\mathcal{P}_\text{top-K}^{(i)}, \mathcal{P}_{\text{GT}}) +(1-\lambda) \cdot \frac{\sum_{t=1}^{T_i} \mathbb{I}(t \in \mathcal{T}_{\text{effective}}^{(i)})}{T_i}
\end{equation}

where $\mathcal{P}_\text{top-K}^{(i)}$ and $\mathcal{P}_{\text{GT}}$ denote the recalled post set and ground-truth target posts, $T_i$ is the total tool call count, and $\mathbb{I}(\cdot)$ indicates effective tool invocations. A tool call $t$ is recognized as a good call ($t \in \mathcal{T}_{\text{effective}}^{(i)}$) if it reduces the candidate post pool size while retaining all target posts within the reduced set, whereas unhelpful or destructive operations are classified as bad calls.

At the micro level, to resolve the credit assignment dilemma, SocialPO augments Group Relative Policy Optimization (GRPO) \cite{shao2024deepseekmathpushinglimitsmathematical} with localized token-level KL supervision. When a rollout fails to achieve a full F1-Score, SocialPO identifies the first bad call and isolates it along with all subsequent steps into $W_i$. The overall SocialPO objective $\mathcal{J}(\theta)$ is defined as:

\begin{equation}
\begin{aligned}
&\mathcal{J}(\theta) = \mathbb{E}_{q, o_i \sim \pi_{\theta_{\text{old}}}} \frac{1}{G} \sum_{i=1}^G \biggl[  \mathcal{L}_{\text{GRPO}}^i(\theta) + 
 \mathbf{1}[R_{\text{f1}}^i < 1] \cdot
 \\ & \frac{\lambda_{kl}}{|W_i|} \sum_{t \in W_i} D_{KL} \left[ \pi_\theta(\cdot \mid q, o_{i,<t}) \parallel \pi_{teacher}(\cdot \mid q, o_{i,<t}) \right] \biggr]
\end{aligned}
\end{equation}

where $\mathcal{L}_{\text{GRPO}}^i(\theta)$ represents the standard trajectory-level GRPO objective driven by the group-relative advantage $\hat{A}_i$, and $\pi_{\text{teacher}}$ denotes the teacher model, which provides reference supervision to enforce dense KL regularization exclusively over the identified erroneous step $W_i$. Through this formulation, SocialPO reinforces high-efficiency macro decisions and rectifies micro-level reasoning missteps.

% To maximize the effectiveness of policy optimization, we construct a challenging training set specifically filtered for policy boundary exploration. Specifically, we focus on hard instances where the initial model fails to find a correct answer under the tightest pass@1 setting, but successfully resolves the task under relaxed pass@5 constraints. Furthermore, we intentionally synthesize complex multi-intent queries that mandate long-horizon search trajectories across compound social constraints. Training on these hard samples ensures that the model operates at its reasoning frontier, pushing the boundaries of trajectory efficiency and tool composition.

% 提示：请确保在导言区（Preamble）包含：\usepackage[table]{xcolor}
\begin{table*}[t]
\centering
\small
\renewcommand{\arraystretch}{1.15}
\setlength{\tabcolsep}{4pt}
\begin{tabular}{c|l|c|cccc|cccc|ccc}
\toprule
\multirow{2}{*}{\textbf{Type}} & \multirow{2}{*}{\textbf{Model}} & \multirow{2}{*}{\textbf{Size}} & \multicolumn{4}{c|}{\textbf{Easy Set}} & \multicolumn{4}{c|}{\textbf{Hard Set}} & \multicolumn{3}{c}{\textbf{Avg. Overall}} \\
\cmidrule(lr){4-7} \cmidrule(lr){8-11} \cmidrule(lr){12-14}
& & & \textbf{SR} & \textbf{Prec.} & \textbf{Recall} & \textbf{EM} & \textbf{SR} & \textbf{Prec.} & \textbf{Recall} & \textbf{EM} & \textbf{Prec.} & \textbf{Recall} & \textbf{EM} \\
\hline
\multirow{4}{*}{\textbf{API-based}} 
& Claude-4.5-Sonnet & N/A & 0.999 & 0.841 & 0.835 & 0.809 & 0.999 & 0.810 & 0.812 & 0.766 & 0.826 & 0.824 & 0.788 \\
& Gemini-2.5-Flash & N/A & 0.913 & 0.732 & 0.747 & 0.684 & 0.868 & 0.635 & 0.627 & 0.609 & 0.684 & 0.687 & 0.647 \\
& Qwen3.5-Plus & N/A & 0.997 & 0.904 & 0.891 & 0.834 & 0.993 & 0.881 & 0.874 & 0.853 & 0.893 & 0.883 & 0.844 \\
& GPT-5 & N/A & 0.998 & 0.912 & 0.921 & 0.892 & 0.995 & 0.884 & 0.887 & 0.848 & 0.898 & 0.904 & 0.870 \\
\hline
\multirow{6}{*}{\textbf{Open-weights}}
& DeepSeek-V4-Flash & 284B & 0.999 & 0.820 & 0.819 & 0.786 &  0.985 & 0.729 & 0.714 & 0.687 & 0.775 & 0.767 &  0.737\\
& Kimi-2.5 & 744B & 0.948 & 0.879 & 0.881 & 0.812 & 0.914 & 0.793 & 0.794 & 0.726 & 0.836 & 0.838 & 0.769 \\

& Qwen3.5-9B & 9B & 0.965 & 0.812 & 0.808 & 0.763 & 0.882 & 0.754 & 0.748 & 0.701 & 0.783 & 0.778 & 0.732 \\
& Qwen3.5-35B-A3B & 35B & 0.988 & 0.855 & 0.857 & 0.818 & 0.918 & 0.811 & 0.804 & 0.780 & 0.833 & 0.831 & 0.799 \\
& Qwen3.5-397B-A17B & 397B & 0.997 & 0.890 & 0.892 & 0.861 & 0.982 & 0.859 & 0.856 & 0.832 & 0.875 & 0.874 & 0.847 \\
& GLM-5.2 & 1.04T & 0.999 & 0.899 & 0.893 & 0.862  & 0.998& 0.879 & 0.881 & 0.843& 0.889 & 0.887 & 0.853 \\
\hline
\rowcolor{gray!12}
 & SocialBuddy-9B & 9B & 0.991 & 0.965 & 0.966 & 0.943 & 0.976 & 0.942 & 0.946 & 0.926 & 0.954 & 0.956 & 0.935 \\
\rowcolor{gray!12}
\multirow{-2}{*}{\textbf{Ours}} & SocialBuddy-35B & 35B & \textbf{0.999} & \textbf{0.974} & \textbf{0.975} & \textbf{0.956} & \textbf{0.996} & \textbf{0.960} & \textbf{0.963} & \textbf{0.943} & \textbf{0.967} & \textbf{0.969} & \textbf{0.950} \\
\bottomrule
\end{tabular}
\caption{Performance comparison between SocialBuddy and baseline models on the Easy and Hard subsets of the SocialSearch benchmark. (Note: Pre. denotes Precision.)}
\label{tab1}
\end{table*}

\subsection{SocialSearch Benchmark}
We construct the SocialSearch Benchmark for comprehensively evaluating SocialBuddy. Constructing real-world social search benchmarks presents an inherent privacy dilemma, where authentic personal social circles and private post histories cannot be publicly released. To respect user privacy while realistically simulating social search environments, we synthesize a dedicated ego-social circle for each of 50 recruited users. Volunteers first familiarize themselves with their assigned post environments and formulate 100 multi-intent search queries each. After filtering out invalid entries, a total of 4,893 query instances were retained.

To establish the target post set $\mathcal{P}_{\text{GT}}$, we adopt a two-stage annotation protocol. A teacher model first executes full-trajectory rollouts across the social circle to retrieve candidate posts satisfying all conditions except the tool-final constraint, and subsequently ranks them for selection. The users then manually evaluate these candidate posts and select at most $K$ relevant posts ($K=5$) as the gold standard. Furthermore, based on query intent complexity and constraint depth, we categorize the benchmark into Easy and Hard subsets to enable fine-grained performance analysis across varying difficulty levels.

\section{Experiment}

\subsection{Experimental Setup}

\subsubsection{DataSet}
Our dataset comprises two partitions tailored for SFT and RL stages. For SFT training, we construct a base dataset of 50K low-complexity trajectories with only two to three intent constraints. Training on these elementary samples guides the model to quickly learn fundamental tool invocation protocols and structured response generation. To maximize policy exploration, we construct a 20K challenging RL dataset. Specifically, we extract 8,745 hard instances failing under strict $\text{pass}$@$1$ but succeeding under $\text{pass}$@$5$ evaluation. We further synthesize 11K complex multi-intent queries with at least three intent constraints. Training on these hard samples actively pushes the model's capability in complex social search task.

\subsubsection{Implementation.}
During the environment and data synthesis phase, Qwen-3.5-plus \cite{qwen35blog}, paired with customized prompts, is utilized as the generation engine for content synthesis. We develop SocialBuddy in two parameter sizes, SocialBuddy-9B and SocialBuddy-35B, built upon Qwen-3.5-9B and Qwen-3.5-35B-A3B base models, respectively. For the SFT stage, the model is trained for $1$ epoch on the SFT dataset. In the SocialPO optimization phase, the group size $G$ is set to $8$, the reward hyperparameter $\lambda$ is set to $0.8$, and the model is trained for $1$ epoch on the RL dataset. At the micro level, the teacher model is trained from Qwen-3.5-122B-A10B through sequential SFT and GRPO on our dataset. The KL regularization hyperparameter $\lambda_{\text{kl}}$ is set to $0.2$, with a token vocabulary size of $64$. For training SocialBuddy-35B via SocialPO, we deploy 32 H20 GPUs across four nodes for policy training alongside 8 H20 GPUs on a single node hosting the teacher model for inference, consuming over 2,000 GPU hours in total.

\begin{figure}[t]
\begin{center}
 \includegraphics[width=\columnwidth]{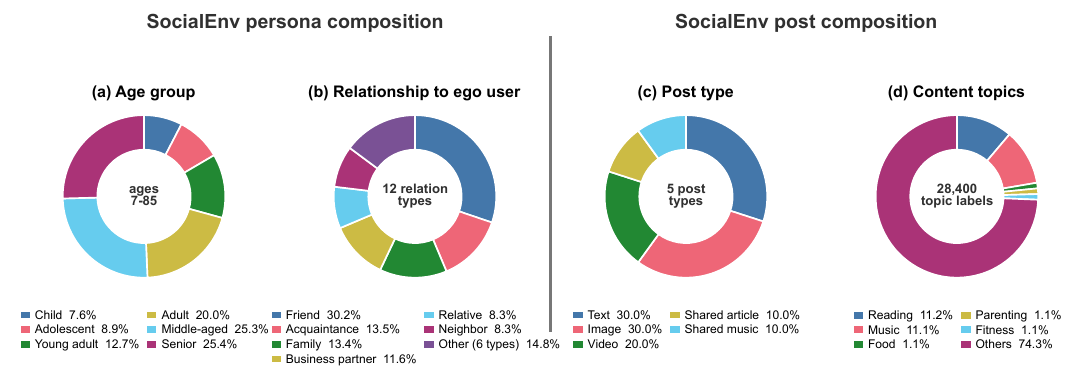}
\end{center}
\caption{Data composition of personas, social relations, post types, and topics in SocialEnv.}

% \textcolor{red}{The complete relationship and topic taxonomies are reported in the supplement.}
\label{socialenv_statistics}
\end{figure}

\begin{table}[t]
\centering
\small
\renewcommand{\arraystretch}{1.15}
\setlength{\tabcolsep}{5pt}
\begin{tabular}{l|ccc}
\toprule
\textbf{Method / Variant} & \textbf{Prec.} & \textbf{Recall} & \textbf{EM} \\
\hline
Qwen3.5-9B (Base) & 0.783 & 0.778 & 0.732 \\
+ SFT & 0.865 & 0.868 & 0.835 \\
+ OPD & 0.921 & 0.924 & 0.909 \\
+ RL (w/ F1-Score Reward only) & 0.892 & 0.905 & 0.878 \\
+ RL (w/ Macro Composite Reward) & 0.914 & 0.921 & 0.894 \\
+ RL (w/ Macro Reward+OPD) & 0.941 & 0.943 & 0.921 \\
\rowcolor{gray!12}
Teacher Model & 0.984 & 0.975 & 0.963 \\
\hline
\textbf{SocialBuddy-9B (Full Method)} & \textbf{0.954} & \textbf{0.956} & \textbf{0.935} \\
\bottomrule
\end{tabular}
\caption{Ablation Study on Key Components of SocialPO.}
\label{tab2}
\end{table}

\subsubsection{Baselines and Evaluation Metrics}
We select 10 representative models for comparison, consisting of 4 proprietary and 6 open-weights baselines, including GPT-5 \cite{singh2026openaigpt5card}, Claude-4.5-Sonnet \cite{anthropic2025claudesonnet45}, Gemini-2.5-Flash \cite{geminiteam2025gemini25}, Qwen3.5-Plus \cite{qwen35blog}, GLM-5.2 \cite{glm5team2026glm5vibecodingagentic}, Kimi-2.5 \cite{kimiteam2026kimik25}, DeepSeek-V4-Flash \cite{deepseekai2026deepseekv4}, Qwen3.5-9B, Qwen3.5-35B-A3B, and Qwen3.5-397B-A17B. To evaluate performance, we measure execution quality across two core dimensions. Specifically, we calculate the task success rate ($\text{SR}$) as the proportion of instances where the model successfully executes the interaction and yields a valid final response. For post retrieval performance, we compare the searched post set $\mathcal{P}_{\text{searched}}$ against the ground-truth target posts $\mathcal{P}_{\text{GT}}$, reporting macro-averaged Precision and Recall across all samples, along with the Exact Match rate ($\text{EM}$) that requires $\mathcal{P}_{\text{searched}} = \mathcal{P}_{\text{GT}}$.

\begin{figure}[t]
\begin{center}
 \includegraphics[width=\columnwidth]{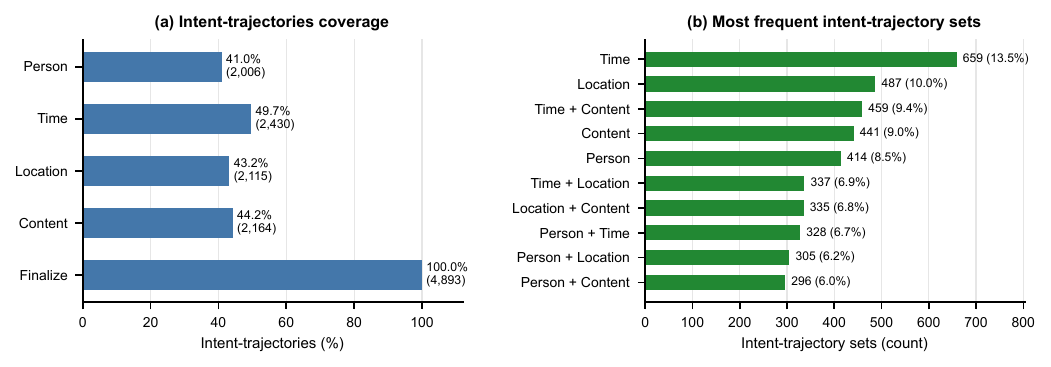}
\end{center}
\caption{Query intent characteristics in the SocialSearch benchmark.
}
\label{socialsearch_benchmark_statistics}
\end{figure} 

\begin{figure*}[t]
\begin{center}
 \includegraphics[width=\linewidth]{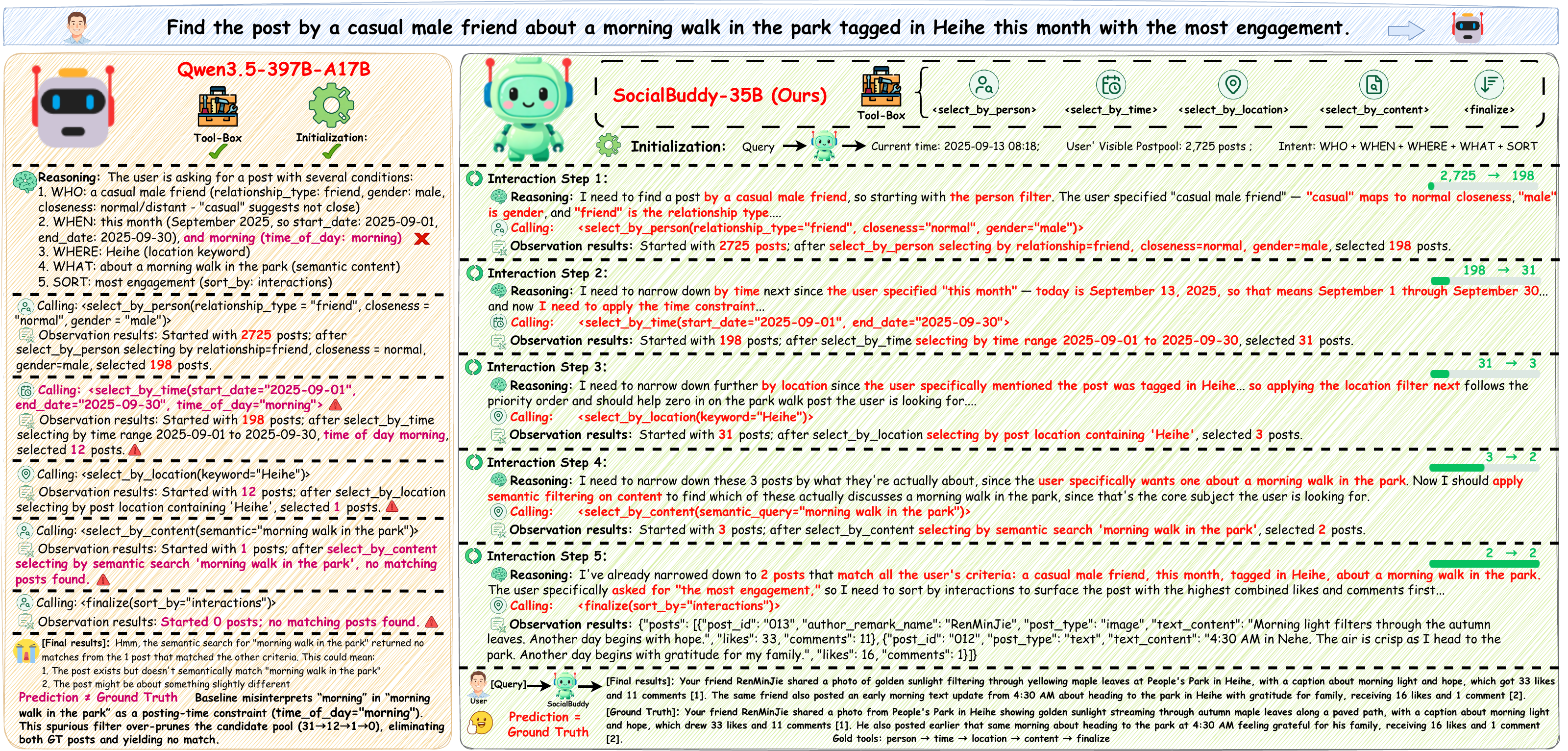}
\end{center}
\caption{Case study comparing Qwen3.5-397B-A17B and SocialBuddy-35B under complex multi-constraint social search.}
% \textcolor{red}{the complete system prompt will be provided in the appendix and released code}.
\label{fig3}
\end{figure*}

\subsection{Environment and Benchmark Analysis}
As shown in Figure~\ref{socialenv_statistics}, SocialEnv simulates a comprehensive social ecosystem comprising 1,000 ego-centric circles, encompassing nearly 200,000 personas and 10 million posts. Personas span ages 7--85 and cover 12 relationship categories relative to ego users. Posts encompass five common social types (text, image, video, shared articles, and music) across 28,400 distinct topic labels.

The SocialSearch benchmark comprises 4,893 validated queries, split into 2,001 easy and 2,892 hard tasks. As shown in Figure~\ref{socialsearch_benchmark_statistics}(a), queries broadly cover key dimensions including person, time, location, and content. Figure~\ref{socialsearch_benchmark_statistics}(b) details the top-10 intent combinations, highlighting that multi-attribute queries dominate the benchmark and reflect realistic compositional complexity.

\subsection{Overall Performance}
The overall performance comparison results on the SocialSearch benchmark are presented in Table~\ref{tab1} and a case study of model capabilities under complex multi-constraint scenarios is illustrated in Figure~\ref{fig3}. Based on these results, we highlight some findings: 1) Our proposed SocialBuddy framework achieves superior performance across all evaluation metrics, outperforming both open-weights and proprietary baselines with significantly fewer parameters regardless of whether evaluated on the Easy Set or the Hard Set. Across both subsets, SocialBuddy-35B consistently leads all competitors, achieving absolute average improvements over the strongest open-weights baseline (GLM-5.2) of 7.8\% in Precision, 8.2\% in Recall, and 9.7\% in Exact Match (EM). It also outperforms GPT-5, the leading proprietary API model, by 6.9\% in Precision, 6.5\% in Recall, and 8.0\% in EM on average. Even our smaller SocialBuddy-9B variant consistently exceeds other competitive baselines on both sets. 2) Our proposed method yields substantial performance gains over the standard base models, demonstrating the core value of our adaptation framework. Compared to the base Qwen3.5-9B model, SocialBuddy-9B elevates average Precision from 0.783 to 0.954 (+17.1\%), Recall from 0.778 to 0.956 (+17.8\%), and EM from 0.732 to 0.935 (+20.3\%). Similarly, SocialBuddy-35B boosts the average performance of Qwen3.5-35B-A3B by 13.4\% in Precision, 13.8\% in Recall, and 15.1\% in EM. These consistent jumps confirm that our method effectively equips general-purpose language models with specialized capabilities for complex social search tasks. 3) Models optimized with our framework achieve near-saturated Task Success Rates, confirming that our method effectively eliminates tool-use execution errors and ensures robust task completion in complex social search scenarios. 4) Case Study further verifies SocialBuddy's superior reasoning under complex multi-constraint scenarios. In a representative five-dimensional query (Figure~\ref{fig3}), SocialBuddy-35B accurately retrieves and ranks target posts by harmonizing semantic and engagement constraints, whereas Qwen3.5-397B-A17B over-constrains the temporal window and returns no results. 

Moreover, we conducted user experience trials to validate SocialBuddy's strong behavioral consistency and practical utility (detailed in Appendix).

% \subsection{Case Study}
% Figure~\ref{fig3} illustrates how intent attribution
% determines the success of candidate selection.
% The query jointly specifies authorial scope, temporal grounding, spatial provenance, content relevance, and an interaction-based presentation objective over $2{,}725$ visible posts. SocialBuddy-35B correctly grounds ``this month,'' maps the author description to structured attributes, and preserves ``morning walk in the park'' for semantic selection, yielding the trajectory
% $2{,}725 \rightarrow 198 \rightarrow 31 \rightarrow 3 \rightarrow 2$. The \textbf{finalize} tool then orders the two relevant posts by 44 and 17 interactions, matching the ground truth. In contrast, Qwen3.5-397B-A17B imposes the unsupported predicate \textbf{time\_of\_day=morning}, shrinking 198 candidates to 12 and ultimately to zero. This comparison demonstrates that correct intent attribution and operator
% boundaries prevent irreversible over-filtering.

\subsection{Ablation Study}
To evaluate key components in SocialPO, we conduct ablation experiments on SocialBuddy-9B (Table~\ref{tab2}). SFT provides essential tool alignment, raising EM from 0.732 to 0.835 over the base model. Among RL variants built on SFT, macro composite reward outperforms the outcome-only F1 reward (0.894 vs. 0.878 EM) by mitigating reward sparsity. While applying standard OPD or combining it with macro rewards boosts performance to 0.909 and 0.921 EM respectively, we find that full-trajectory preference supervision is less optimal than targeted error supervision. By isolating failed dimensions and specifically penalizing execution missteps, our full SocialBuddy-9B framework achieves the highest performance at 0.935 EM, closely approaching the Teacher Model.
%\subsection{Case Analysis}

\section{Conclusion}
In this paper, we propose SocialBuddy, the first agentic search framework tailored for social scenarios. To capture the characteristics of real-world social platforms, we construct SocialEnv, the first large-scale simulated social environment for socail search task. Furthermore, we introduce SocialPO, a hybrid-granularity policy optimization framework designed to simultaneously reinforce valid decisions and rectify localized errors. Finally, to systematically assess model capabilities, we establish the SocialSearch benchmark covering diverse intent complexity and search trajectories. Extensive experimental results demonstrate that SocialBuddy-35B surpasses significantly larger frontier LLMs, effectively extending the capabilities of conventional agentic search into the realm of daily social intelligence.

\bibliography{aaai2027}

\clearpage % Draft layout only; remove for the AAAI final version.

\section{Appendix \uppercase\expandafter{\romannumeral 1}: User Experience Evaluation}
%\subsection{}
While evaluations on SocialBenchmark demonstrate strong performance, its underlying search environment remains semi-synthetic with publicly sourced posts, despite incorporating real-world user queries. In real-world applications, models must handle users' implicit search intents directly over highly dynamic, personal social feeds. To evaluate SocialBuddy's real-world generalization, we conducted an user study in genuine private social environments.

Specifically, we recruited 50 experienced social platform users as volunteers. With explicit informed consent and strict privacy preservation protocols, we deployed four target models including GLM-5.2, Qwen3.5-397B-A17B, Qwen3.5-35B-A3B, and SocialBuddy-35B to interface with each participant's real personal social feed.

Each participant performed 20 ad hoc natural language searches against their own personal social history, yielding a total of 1,000 comparative evaluation trials. To eliminate evaluation bias, a double-blind protocol was enforced: the responses from the four models were anonymized and presented in a randomized order.

Model capabilities were evaluated across two aggregate metrics:
\begin{itemize}
    \item \textbf{Hallucination Rate:} A model response is marked as hallucinated if it either returns non-existent/fabricated posts or contains descriptions/citations that contradict the actual retrieved content. The overall Hallucination Rate is computed as the percentage of such hallucinated responses across all $1{,}000$ evaluation trials.
    \item \textbf{Top-$k$ Preference Rate:} The proportion of evaluation trials in which a model's response was ranked within the top-$k$ model.
\end{itemize}

% 过渡句
The user experience evaluation results for the four models across the real-world trials are summarized in Table~\ref{tab:user_study_results}. Notably, all four models exhibit remarkably low and comparable Hallucination Rates ($\le 1.6\%$), with GLM-5.2 reaching 0.8\%. This outcome directly validates our core design philosophy: rather than overwhelming the LLM with excessive raw posts that degrade context clarity, our framework only feeds the top-$k$ retrieved posts after the final $Tool_{fin}$ execution phase. Since modern foundation models already possess strong summarization and synthesis capabilities, removing context clutter allows them to faithfully restrict responses to the retrieved evidence. However, when evaluating user satisfaction, \textbf{SocialBuddy-35B} demonstrates overwhelming dominance. It secures a 64.2\% Top-1 Preference Rate, which is nearly triple that of the strongest baseline (Qwen3.5-397B-A17B at 21.4\%), and achieves over 90\% in Top-2 Satisfaction Rate. This proves that our model's targeted search strategy and synthesized outputs seamlessly align with real-world user intent and personalized needs, a practical outcome that stands in tight alignment with our offline evaluation results on the Social Search Benchmark.

\begin{table}[h]
\centering
\resizebox{\linewidth}{!}{%
\begin{tabular}{lccc}
\toprule
\textbf{Model} & \textbf{HR (\%)} $\downarrow$ & \textbf{Top-1 (\%)} $\uparrow$ & \textbf{Top-2 (\%)} $\uparrow$ \\
\midrule
Qwen3.5-35B-A3B & 1.6 & 8.3 & 23.8 \\
GLM-5.2 & \textbf{0.8} & 13.1 & 38.2 \\
Qwen3.5-397B-A17B & 1.3 & 21.4 & 46.5 \\
\textbf{SocialBuddy-35B (Ours)} & 1.4 & \textbf{64.2} & \textbf{91.5} \\
\bottomrule
\end{tabular}%
}
\caption{User experience evaluation results across $1{,}000$ comparative trials in real-world personal social environments. HR, Top-1, and Top-2 denote Hallucination Rate, Top-1 Preference Rate, and Top-2 Preference Rate, respectively.}
\label{tab:user_study_results}
\end{table}

\section{Appendix \uppercase\expandafter{\romannumeral 2}: Tool Design and Interaction Protocol}

This section specifies the implementation contract underlying the five-tool action space introduced in Section 3 of the main text. It details the structural design, parameter definitions, and execution flow of the interaction protocol.

\subsection{Design Principles}
\label{app:stateful-pool}
Social queries may jointly constrain author, time, location, content, and result order. SocialBuddy maintains a stateful candidate pool and factorizes these dimensions into four selectors followed by a terminal ranking tool. For retrieval dimension $a$, let $\phi_a(p;c_a)\in\{0,1\}$ denote whether post $p$ satisfies criterion $c_a$, define $S_a(c_a)=\{p\mid\phi_a(p;c_a)=1\}$ as its satisfaction set, and let $\Omega_a$ denote the corresponding selector. A successful tool execution updates the pool by
\begin{equation}
\begin{aligned}
P_{k+1}
&=\Omega_{a_k}(P_k;c_{a_k})\\
&=\{p\in P_k\mid\phi_{a_k}(p;c_{a_k})=1\}
=P_k\cap S_{a_k}(c_{a_k}).
\end{aligned}
\label{eq:appendix-pool-transition}
\end{equation}

Hence, evaluated constraints compose conjunctively and the pool contracts monotonically: $\Omega_a(P;c_a)\subseteq P$. The policy invokes only query-supported dimensions, preventing unexpressed facets from becoming retrieval predicates.

The factorization also induces a cost-aware execution order. Author, temporal, and POI predicates use inexpensive in-memory comparisons, whereas open-vocabulary relevance requires batched model inference whose cost grows with the surviving pool. SocialBuddy therefore applies all requested structured selectors before content selection, and evaluates exact keyword matches before dense semantic scoring within the content tool. A single terminal call follows the completed selection sequence. This coarse-to-fine schedule reduces semantic-inference latency and preserves facet-level traceability.

\subsection{Tool Roles and Semantic Boundaries}
\label{app:tool-interface}
Table~\ref{tab:tool-interface} maps each tool to a distinct retrieval responsibility. The first three tools operate on structured author or post metadata. \texttt{select\_by\_content} applies an optional modality predicate before open-vocabulary relevance assessment, while \texttt{finalize} orders, truncates, and projects the surviving evidence without introducing an additional relevance condition. Because all selectors update the same pool, every earlier constraint remains active in subsequent steps. Intermediate observations expose the applied condition and candidate-count transition; ordered post records are disclosed only by the terminal call.

\subsubsection{Selection Grounded in Explicit Attributes}
Tables~\ref{tab:person-parameters} and~\ref{tab:time-parameters} specify the two priority-1 selectors. \texttt{select\_by\_person} conjunctively matches all supplied relationship, closeness, gender, and age-cohort attributes. Name matching removes whitespace, lowercases the query, and tests it as a substring of either the author's nickname or the querying user's remark name. The four public age cohorts map to finer SocialEnv categories; no numeric age is inferred during retrieval.

\texttt{select\_by\_time} evaluates inclusive date bounds, recorded weekdays, structured festival annotations, and fixed half-open time-of-day buckets over publication metadata. The policy resolves relative expressions against the query reference time before invocation, so the tool receives absolute dates.

The priority-2 \texttt{select\_by\_location} contract appears in Table~\ref{tab:remaining-parameters}. It normalizes case and apostrophe variations, strips standard administrative suffixes across different granularity levels (e.g., municipal, provincial, and district designations), and then executes substring matching strictly against \texttt{poi\_city} or \texttt{poi\_name}.

% \FloatBarrier
\subsubsection{Hybrid Content Selection}
\texttt{select\_by\_content} combines a deterministic modality predicate with an auxiliary semantic selector (Table~\ref{tab:remaining-parameters}). When \texttt{post\_type} is supplied, exact enum matching precedes semantic inference. The surviving records are serialized as numbered summaries containing authored text, media or share descriptions, modality, POI, festival annotation, media counts, and engagement counts. Identifiers, author attributes, raw timestamps, weekdays, and generated topic labels are excluded. A record is retained only when the query matches its principal communicative subject rather than an incidental mention or background element.

Semantic selection represents the sole model-based filtering stage, powered by the Qwen-3.5-397B-A17B model. To enforce bounded concurrency, the runtime evaluates candidate posts in batches of up to 100, requiring $\lceil m/100\rceil$ forward passes for a pool of size $m$. Consequently, earlier structured predicates directly minimize both inference latency and execution cost, while an empty return simply indicates no semantic match.

\subsubsection{Terminal Ranking and Public Output}
\texttt{finalize} is the terminal action and does not alter relevance membership. It orders survivors by \texttt{likes}, \texttt{comments}, their sum (\texttt{interactions}), or the default importance score
\begin{equation}
s_i=0.3r_i+0.4l_i+0.3c_i,
\label{eq:importance-score}
\end{equation}
where $r_i=(t_i-\min_jt_j)/(\max_jt_j-\min_jt_j)$ is candidate-local min--max recency (or $0.5$ when all timestamps coincide), while $l_i=\mathrm{likes}_i/\max_j\mathrm{likes}_j$ and $c_i=\mathrm{comments}_i/\max_j\mathrm{comments}_j$ (zero when the respective maximum is zero). Equal primary scores are resolved deterministically by ascending publication time and then post identifier.

As detailed in Table~\ref{tab:remaining-parameters}, the output is capped at five records and exposes no agent-controlled count argument. The public projection retains content, publication time, POI, engagement counts, and minimal contact descriptors needed for attribution, while excluding author age, gender, residence, and internal annotations. The returned order defines the citation index $[n]$ used in the natural-language response.

\subsection{Interaction, Feedback, and Failure Semantics}
\label{app:invocation-protocol}
The policy first decomposes the query, then applies one query-supported retrieval dimension per action, and finally synthesizes a response from terminal evidence. Intermediate feedback is restricted to the tool identity, applied condition, and candidate-count transition rather than repeatedly disclosing the active records. Only \texttt{finalize} returns the ordered public evidence used for answer composition and citation.

\subsection{Trajectory-Level Evidence}
\label{app:tool-trajectories}
Beyond the trajectory in the main paper, Tables~\ref{tab:trajectory-example}--\ref{tab:trajectory-c574} provide eight additional Hard-set executions. Five fully aligned trajectories cover author cohorts, relative and event-based temporal grounding, modality constraints, and observable multi-result ranking. Three diagnostic trajectories isolate distinct errors: over-decomposing a joint content phrase into a location predicate, treating a depicted child as an author-age constraint, and citing an empty terminal result.

The target policy decomposes each query into its expressed WHO, WHEN, WHERE, WHAT, and SORT facets, applies low-cost structured predicates before semantic content matching, and terminates with a single ranking call. Each table retains the user query, reference time, decision-relevant rationale, normalized arguments, candidate-count transitions, terminal public evidence, and generated response. Repetitive rationale is abridged, while intermediate observations omit record identifiers.

\subsection{Prompt Template and Public Action Schema}
\label{app:prompt-template}
The deployed policy prompt has three separable components: a task specification that defines the social-search setting, a decision policy that maps expressed intent dimensions to an ordered tool sequence, and a serialization protocol that makes tool calls and citations machine-checkable. Tables~\ref{tab:system-prompt-policy} and~\ref{tab:system-prompt-protocol} present these policy-bearing instructions in a normalized layout. The tool schemas referenced by the prompt are factored into Tables~\ref{tab:person-parameters}--\ref{tab:remaining-parameters}, allowing the semantics of every public argument to be examined independently of its serialized representation.

\FloatBarrier
\section{Appendix \uppercase\expandafter{\romannumeral 3} Persona Composition and Examples}
\label{app:persona-composition}
\paragraph{Construction protocol.}
SoMe (Xue et al., 2026) constructs social-agent tasks from real-world public posts, author profiles, and external reports. Images are normalized through captions and OCR, video speech is transcribed, and the resulting multimodal evidence is exposed through dedicated tools. Depending on the task, queries and targets are constructed automatically, annotated through a human--LLM workflow with professional verification, or inherited from the source datasets. This pipeline organizes and verifies heterogeneous observed records; it does not synthesize persona-conditioned social histories.

SocialEnv instead synthesizes privacy-preserving, persona-conditioned histories. It first instantiates a structured user representation spanning demographics, education, occupation, household context, routines, values, interests, and communication style. The representation is partitioned into public author attributes and private generation context: the complete profile conditions post synthesis, whereas social search exposes only authorized public attributes and observable records. The generator combines persona context with timestamps sampled from human activity patterns, topic and situational cues, and style controls to produce text, image, video, music-share, and article-share records. The resulting histories are placed in ego-centric first-degree circles that define the environment's visibility boundary.

\paragraph{Persona representation.}
Tables~\ref{tab:persona-showcase}--\ref{tab:persona-practical-bloom} instantiate the representation with five profiles spanning distinct life stages, regions, occupations, household roles, interests, and communication styles. Each table preserves the joint configuration of demographic background, family and work context, routine, disposition, narrative anchors, and ego-network relation because these dimensions jointly condition the associated social history. Long narrative fields are compressed into their constituent events and behavioral cues.

A persona record contains 14 top-level blocks and 37 leaf fields. The tables retain age and life stage, gender and region, education, family composition, occupation, housing and mobility, daily routine, values, attitude and coping, interests, six expression controls, narrative grounding, relationship type, and tie strength. Values encode stable priorities, attitude specifies the profile's interpretive stance, and coping describes its response strategy; these fields therefore remain distinct. The examples omit the internal profile index and raw contact-specific nickname and remark strings. Their operational roles in name matching, authorship, and social linkage are specified in Tables~\ref{tab:person-parameters} and~\ref{tab:persona-record-composition}.

The \texttt{personal\_story} field is represented by compact narrative anchors rather than reproduced as an extended biography. These anchors retain the events through which static attributes acquire behavioral consequences---for example, resolving a server outage, mediating a neighborhood dispute, or rescheduling a delayed shipment. They link structured conditioning variables to recurrent actions without allowing narrative length to dominate the profile comparison.

The five profiles also demonstrate that demographic tags alone are insufficient to specify social behavior. \mbox{GreenThumbChef} combines technical work with rural multigenerational routines; Grandma Li centers late-life community participation and intergenerational care; Steady Hearth couples media supervision with responsibility for a young granddaughter; Harmony Home connects ethical healthcare communication with cross-generational care; and Practical Bloom combines urban sales work, household logistics, gardening, and child care. Expression controls provide an additional axis governing topic choice and linguistic realization.

Table~\ref{tab:persona-record-composition} gives the corresponding visibility contract. Public author attributes support person-level retrieval, whereas private context conditions synthesis but is not available as a selector predicate. A post may therefore express a routine, value, or household circumstance without making the underlying profile field directly accessible to the search agent. This separation preserves persona-conditioned behavioral coherence while maintaining the information boundary defined in the main paper.

\FloatBarrier
\section{Appendix \uppercase\expandafter{\romannumeral 4} Post Composition and Examples}
\label{app:post-composition}
\paragraph{Post representation.}
As described in the main paper, SocialEnv contains 10 million posts synthesized for 200K user profiles and organized into 1,000 ego-centric social circles. Each structured record belongs to one of five modalities---text, image, video, music share, or article share---and stores authored content together with timestamp and topic metadata. Records may additionally include post-level POI provenance and simulated like and comment counts. Image and video records retain textual descriptions of their media payloads, enabling a common structured retrieval interface across modalities without reconstructing visual content.

Tables~\ref{tab:post-showcase}--\ref{tab:post-showcase-steady-hearth} report fifteen records for each of \mbox{GreenThumbChef}, Grandma Li, and Steady Hearth, yielding 45 examples with complete coverage of the five supported modalities. Each row includes the source-local post identifier, modality, timestamp, weekday, festival annotation, topic, optional post-level POI, engagement, and applicable modality-specific payload. Authored text is reproduced verbatim after emoji removal. Potentially identifying POI names are coarsened or withheld, Chinese shared-item strings are translated into English, and long media descriptions and shared-item metadata are abridged without changing the reported evidence.

\paragraph{Persona--post consistency.}
\mbox{GreenThumbChef}'s records couple information-systems work with rural family routines. The agricultural-drone article, HRIS workstation sequence, and work--family music share express her technical background across textual, visual, and musical content. Dawn meal preparation, e-bike school trips, home maintenance, gardening, and pauses during the commute expose household planning, child care, and everyday rural life. Concise observation and brief reflection remain consistent with the specified Life Logger style across modalities.

Grandma Li's records realize the Family Elder profile through Tai Chi, traditional meals, family memory, and reciprocal intergenerational learning. Recurrent park practice supplies a stable routine without collapsing the history into a single topic: river memories and music recall her late spouse; family meals and a pipa performance foreground multigenerational continuity; and the tablet and television sequences show her adult grandson supporting engagement with new media. The college-guide article further expresses the careful guidance associated with her community-service background. Warm and grateful phrasing remains stable across all five modalities.

Steady Hearth's records connect professional responsibility with her role as a family anchor. Team-safety advice and the late-night office video express a deliberative supervisory stance, while the overnight fever watch, kindergarten pickup, festival cooking, dumpling preparation, and late snack with her son make multigenerational care observable. Her article share interprets employment stability through concern for her adult son's family. Occupation, household role, and expression style consequently remain identifiable across heterogeneous content and the ten-month interval covered by the records.

These correspondences assess conditional consistency rather than requiring every post to express every persona attribute. Topic, time, and modality vary, while recurrent routines, concerns, and expression styles remain recoverable at the history level.

Table~\ref{tab:post-record-composition} summarizes the post schema and maps each field group to its retrieval or presentation role. It distinguishes author residence from post-level POI provenance, observable content from generated topic annotations, and semantic evidence from engagement signals used only for terminal ranking.

\begin{table*}[!t]
\caption{Public action space of SocialBuddy. Each tool realizes a formal operator with an explicit parameterization and evidence boundary. Selection tools intersect their predicates with the active candidate pool; \texttt{finalize} is called exactly once at the end.}
\label{tab:tool-interface}
\centering
\scriptsize
\renewcommand{\arraystretch}{1.10}
\setlength{\tabcolsep}{3pt}
\begin{tabular}{@{}>{\raggedright\arraybackslash}p{0.16\linewidth}|>{\raggedright\arraybackslash}p{0.075\linewidth}|>{\raggedright\arraybackslash}p{0.275\linewidth}|>{\raggedright\arraybackslash}p{0.45\linewidth}@{}}
\toprule
\textbf{Tool} & \textbf{Operator} & \textbf{Parameters} & \textbf{Responsibility and evidence boundary} \\
\midrule
\texttt{select\_by\_person}
& $\Omega_{\mathrm{author}}$
& \texttt{name}; \texttt{relationship\_type}; \texttt{closeness}; \texttt{gender}; \texttt{age\_group}
& Conjunctively filters identity, authorized social-graph attributes, and demographic cohorts. It does not infer authorship from post content or residence. \\
\rowcolor{AppStripe}
\texttt{select\_by\_time}
& $\Omega_{\mathrm{temporal}}$
& \texttt{start\_date}; \texttt{end\_date}; \texttt{weekday}; \texttt{festival}; \texttt{time\_of\_day}
& Grounds absolute, periodic, and event-driven constraints in publication metadata rather than textual mentions; date bounds are inclusive. \\
\texttt{select\_by\_location}
& $\Omega_{\mathrm{spatial}}$
& \texttt{keyword}
& Restricts spatial provenance to normalized \texttt{poi\_city} or \texttt{poi\_name} evidence attached to a post, explicitly excluding author residence. \\
\rowcolor{AppStripe}
\texttt{select\_by\_content}
& $\Omega_{\mathrm{content}}$
& \texttt{post\_type}; \texttt{semantic\_query}
& Applies an exact modality gate, then batched semantic assessment of principal-subject relevance over the survivors; private author fields and generated topics are withheld. \\
\texttt{finalize}
& $R_{\rho}$
& \texttt{sort\_by}
& Orders only eligible candidates by the requested objective and returns at most five sanitized records; it neither introduces a new relevance predicate nor re-filters the pool. \\
\bottomrule
\end{tabular}
\end{table*}

\begin{table*}[!t]
\caption{Implementation contract of \texttt{select\_by\_person}. Supplied arguments are conjoined, except that \texttt{name} may match either the normalized nickname or remark name.}
\label{tab:person-parameters}
\centering
\footnotesize
\renewcommand{\arraystretch}{1.03}
\setlength{\tabcolsep}{3.5pt}
\begin{tabular}{@{}>{\raggedright\arraybackslash}p{0.18\linewidth}|>{\raggedright\arraybackslash}p{0.13\linewidth}|>{\raggedright\arraybackslash}p{0.29\linewidth}|>{\raggedright\arraybackslash}p{0.33\linewidth}@{}}
\toprule
\textbf{Argument} & \textbf{Type / status} & \textbf{Accepted input} & \textbf{Matching rule} \\
\midrule
\rowcolor{AppBand}\multicolumn{4}{@{}l}{\textbf{Author metadata selection --- \texttt{select\_by\_person}}}\\
\texttt{name}
& string; optional
& Free-form name or alias
& Remove all whitespace and lowercase; retain an author when the query is a substring of either \texttt{nickname} or \texttt{remark\_name}. \\
\rowcolor{AppStripe}
\texttt{relationship\_}\newline\texttt{type}
& enum; optional
& \texttt{self}, \texttt{family}, \texttt{relative}, \texttt{significant\_other}, \texttt{colleague}, \texttt{college\_classmate}, \texttt{school\_classmate}, \texttt{friend}, \texttt{online\_friend}, \texttt{acquaintance}, \texttt{neighbor}, \texttt{business\_partner}
& Case-normalized exact match on the stored relationship category. \\
\texttt{closeness}
& enum; optional
& \texttt{close}, \texttt{normal}, \texttt{distant}, \texttt{self}
& Case-normalized exact match on tie strength. \\
\rowcolor{AppStripe}
\texttt{gender}
& enum; optional
& \texttt{male}, \texttt{female}
& Case-normalized exact match on profile metadata. \\
\texttt{age\_group}
& enum; optional
& \texttt{child}, \texttt{young}, \texttt{middle}, \texttt{senior}
& Maps respectively to stored categories \texttt{child}; \texttt{adolescent}/\texttt{young\_adult}; \texttt{adult}/\texttt{middle\_aged}; and \texttt{senior}. Public interpretations are 7--12, 13--29, 30--65, and 66+. \\
\bottomrule
\end{tabular}
\smallskip
\parbox{0.95\linewidth}{\footnotesize\textit{Implementation boundary.} The tool reads author metadata only; it does not inspect post content or author residence.}
\end{table*}

\begin{table*}[!t]
\caption{Implementation contract of \texttt{select\_by\_time}. Supplied predicates are conjoined over publication metadata; relative language is resolved by the policy before invocation.}
\label{tab:time-parameters}
\centering
\footnotesize
\renewcommand{\arraystretch}{1.03}
\setlength{\tabcolsep}{3.5pt}
\begin{tabular}{@{}>{\raggedright\arraybackslash}p{0.14\linewidth}|>{\raggedright\arraybackslash}p{0.14\linewidth}|>{\raggedright\arraybackslash}p{0.29\linewidth}|>{\raggedright\arraybackslash}p{0.36\linewidth}@{}}
\toprule
\textbf{Argument} & \textbf{Type / status} & \textbf{Accepted input} & \textbf{Matching rule} \\
\midrule
\rowcolor{AppBand}\multicolumn{4}{@{}l}{\textbf{Publication-time selection --- \texttt{select\_by\_time}}}\\
\texttt{start\_date}
& string; optional
& \texttt{YYYY-MM-DD}
& Inclusive lower calendar-date bound on \texttt{post\_time}. \\
\rowcolor{AppStripe}
\texttt{end\_date}
& string; optional
& \texttt{YYYY-MM-DD}
& Inclusive upper calendar-date bound on \texttt{post\_time}. \\
\texttt{weekday}
& string; optional
& Full name/abbreviation; \texttt{0}--\texttt{6} (Monday=\texttt{0}); \texttt{weekday}; \texttt{weekend}
& Normalize to English weekday names and exactly match the recorded \texttt{weekday} field. \\
\rowcolor{AppStripe}
\texttt{festival}
& enum; optional
& One canonical festival label
& Intersect with the recorded festival list. \texttt{Spring Festival} additionally includes its first-, second-, and third-day labels; no post-text inference is used. \\
\texttt{time\_of\_day}
& enum; optional
& \texttt{morning}, \texttt{noon}, \texttt{afternoon}, \texttt{evening}, \texttt{night}, \texttt{midnight}
& Match half-open hour buckets $[05,11)$, $[11,13)$, $[13,17)$, $[17,21)$, $[21,24)$, and $[00,05)$, respectively. \\
\bottomrule
\end{tabular}
\smallskip
\parbox{0.95\linewidth}{\footnotesize\textit{Canonical festival enum (31 values).} Arbor Day; Army Day; Children's Day; Chinese New Year's Eve; Christmas; Christmas Eve; Cold Food Festival; Double Ninth Festival; Dragon Boat Festival; Father's Day; Halloween; Laba Festival; Labor Day; Lantern Festival; Little New Year; Mid-Autumn Festival; Mother's Day; National Day; New Year's Day; Qingming Festival; Qixi Festival; Singles' Day; Spring Festival; Spring Festival (1st Day); Spring Festival (2nd Day); Spring Festival (3rd Day); Teacher's Day; Valentine's Day; Winter Solstice; Women's Day; Youth Day.}
\end{table*}

\begin{table*}[!t]
\caption{Implementation contracts of the spatial, content, and terminal tools. Gray bands distinguish low-cost structured selection, hybrid content selection, and terminal presentation.}
\label{tab:remaining-parameters}
\centering
\footnotesize
\renewcommand{\arraystretch}{1.16}
\setlength{\tabcolsep}{3.5pt}
\begin{tabular}{@{}>{\raggedright\arraybackslash}p{0.18\linewidth}|>{\raggedright\arraybackslash}p{0.14\linewidth}|>{\raggedright\arraybackslash}p{0.26\linewidth}|>{\raggedright\arraybackslash}p{0.36\linewidth}@{}}
\toprule
\textbf{Tool / argument} & \textbf{Type / status} & \textbf{Accepted input} & \textbf{Implementation and output} \\
\midrule
\rowcolor{AppBand}\multicolumn{4}{@{}l}{\textbf{Spatial selection --- \texttt{select\_by\_location}}}\\
\texttt{keyword}
& string; required
& City or POI phrase
& Normalize case, apostrophes, and common trailing administrative suffixes; substring-match \texttt{poi\_city} or \texttt{poi\_name}. Update pool membership and report before/after counts. \\
\rowcolor{AppBand}\multicolumn{4}{@{}l}{\textbf{Hybrid content selection --- \texttt{select\_by\_content}}}\\
\texttt{post\_type}
& enum; optional
& \texttt{image}, \texttt{text}, \texttt{video}, \texttt{share\_music}, \texttt{share\_article}
& Apply an exact, in-memory modality gate before semantic inference. \\
\rowcolor{AppStripe}
\texttt{semantic\_query}
& string; optional
& Open-vocabulary natural-language criterion
& Select principal-subject matches from numbered post summaries, at most 100 per batch. Author fields, identifiers, raw time, weekday, and generated topics are withheld. \\
\rowcolor{AppBand}\multicolumn{4}{@{}l}{\textbf{Terminal presentation --- \texttt{finalize}}}\\
\texttt{sort\_by}
& enum; required
& \texttt{importance}, \texttt{likes}, \texttt{comments}, \texttt{interactions}
& Sort without re-filtering. Importance uses Equation~\ref{eq:importance-score}; other modes use descending counts. Return at most five slim records; no \texttt{top\_n} argument is exposed. \\
\bottomrule
\end{tabular}
\smallskip
\parbox{0.95\linewidth}{\footnotesize\textit{Operational notes.} Semantic batches use bounded concurrency and, by default, up to three retries after the first attempt. A retry-exhausted batch is reported and contributes no selected item; a wholly unavailable semantic stage is skipped with an explicit warning. Equal ranking scores are resolved by ascending publication time and then post identifier.}
\end{table*}

\begin{table*}[!t]
\caption{Successful trajectory combining named-author, rolling-window, time-of-day, location, text-modality, semantic, and likes constraints (task \texttt{c559\_0010}).}
\label{tab:trajectory-example}
\centering
\footnotesize
\renewcommand{\arraystretch}{1.10}
\setlength{\tabcolsep}{5pt}
\begin{tabular}{@{}>{\raggedright\arraybackslash\sffamily\scriptsize}p{0.13\textwidth}>{\raggedright\arraybackslash}p{0.82\textwidth}@{}}
\toprule
\textbf{Role} & \textbf{Content} \\
\midrule
\rowcolor{AppBand}
User
& \textbf{Query:} show me HongJunYan's text-only posts about reflecting on farming from Burqin in the evening over the last 30 days, sorted by most likes\newline
\textbf{Reference time:} Monday, September 8, 2025, 02:23. \\
Agent $\cdot$ Step 1
& \emph{The query names HongJunYan, so I begin with the person filter.}\newline
\texttt{select\_by\_person(name="HongJunYan")} \\
\rowcolor{AppFill}
Observation
& \textbf{2,041 $\rightarrow$ 23.} The normalized name matched \texttt{remark\_name=HongJunYan}. \\
Agent $\cdot$ Step 2
& \emph{Resolve ``last 30 days'' to August 10 through September 8 and retain evening posts.}\newline
\texttt{select\_by\_time(start\_date="2025-08-10", end\_date="2025-09-08", time\_of\_day="evening")} \\
\rowcolor{AppFill}
Observation
& \textbf{23 $\rightarrow$ 4.} Matched the inclusive date range and evening bucket. \\
Agent $\cdot$ Step 3
& \emph{Enforce the explicit Burqin provenance before inspecting content.}\newline
\texttt{select\_by\_location(keyword="Burqin")} \\
\rowcolor{AppFill}
Observation
& \textbf{4 $\rightarrow$ 2.} Matched post locations containing ``Burqin.'' \\
Agent $\cdot$ Step 4
& \emph{Gate on text modality before evaluating ``reflecting on farming.''}\newline
\texttt{select\_by\_content(post\_type="text", semantic\_query="reflecting on farming")} \\
\rowcolor{AppFill}
Observation
& \textbf{2 $\rightarrow$ 1.} The type gate and semantic selector retained one post. \\
Agent $\cdot$ Step 5
& \emph{``Most likes'' maps to the terminal likes order.}\newline
\texttt{finalize(sort\_by="likes")} \\
\rowcolor{AppFill}
Final observation
& \textbf{1 $\rightarrow$ 1.} HongJunYan; 2025-08-31 20:45:22; text; Burqin County Courtyard; ``My grandson says robots will farm soon. My hands still ache from the old way.''; 3 likes and 1 comment. \\
\rowcolor{AppBand}
Final answer
& Your friend HongJunYan shared a text post from Burqin County Courtyard in the evening, reflecting on how their hands still ache from traditional farming even as their grandson talks about robots taking over the work [1]. \\
\bottomrule
\end{tabular}
\end{table*}

\begin{table*}[!t]
\caption{Successful trajectory combining friend scope, a resolved weekly interval, midnight publication time, video modality, semantic content, and likes ordering (task \texttt{c94\_0063}).}
\label{tab:trajectory-c94}
\centering
\footnotesize
\renewcommand{\arraystretch}{1.10}
\setlength{\tabcolsep}{5pt}
\begin{tabular}{@{}>{\raggedright\arraybackslash\sffamily\scriptsize}p{0.13\textwidth}>{\raggedright\arraybackslash}p{0.82\textwidth}@{}}
\toprule
\textbf{Role} & \textbf{Content} \\
\midrule
\rowcolor{AppBand}
User
& \textbf{Query:} show me the video from friends last week late at night tagged in Weifang about showing off denim jackets with the most likes\newline
\textbf{Reference time:} Friday, July 4, 2025, 12:27. \\
Agent $\cdot$ Step 1
& \emph{Restrict the pool to posts from friends.}\newline
\texttt{select\_by\_person(relationship\_type="friend")} \\
\rowcolor{AppFill}
Observation
& \textbf{1,030 $\rightarrow$ 251.} Relationship metadata matched. \\
Agent $\cdot$ Step 2
& \emph{Resolve last week to June 23--29 and map late night to midnight.}\newline
\texttt{select\_by\_time(start\_date="2025-06-23", end\_date="2025-06-29", time\_of\_day="midnight")} \\
\rowcolor{AppFill}
Observation
& \textbf{251 $\rightarrow$ 11.} The date range and midnight bucket matched. \\
Agent $\cdot$ Step 3
& \emph{Apply the Weifang provenance constraint.}\newline
\texttt{select\_by\_location(keyword="Weifang")} \\
\rowcolor{AppFill}
Observation
& \textbf{11 $\rightarrow$ 1.} One Weifang-tagged post remained. \\
Agent $\cdot$ Step 4
& \emph{Gate on video before checking the denim-jacket subject.}\newline
\texttt{select\_by\_content(post\_type="video", semantic\_query="showing off denim jackets")} \\
\rowcolor{AppFill}
Observation
& \textbf{1 $\rightarrow$ 1.} The modality and semantic criteria matched. \\
Agent $\cdot$ Step 5
& \emph{Apply the requested likes order.}\newline
\texttt{finalize(sort\_by="likes")} \\
\rowcolor{AppFill}
Final observation
& \textbf{1 $\rightarrow$ 1.} HeZhangYing; 2025-06-28 01:09:13; video; Shiqiaozi Industrial Park, Weifang; newly finished denim jackets under factory lights; 0 likes and 0 comments. \\
\rowcolor{AppBand}
Final answer
& Your close friend HeZhangYing shared a late-night video from Shiqiaozi Industrial Park in Weifang, showing off a rack of newly finished denim jackets with perfect stitching under bright factory lights [1]. \\
\bottomrule
\end{tabular}
\end{table*}

\begin{table*}[!t]
\caption{Successful trajectory grounding self-authorship, Spring Festival metadata, POI provenance, semantic content, and comments ordering (task \texttt{c304\_0003}).}
\label{tab:trajectory-c304}
\centering
\footnotesize
\renewcommand{\arraystretch}{1.10}
\setlength{\tabcolsep}{5pt}
\begin{tabular}{@{}>{\raggedright\arraybackslash\sffamily\scriptsize}p{0.13\textwidth}>{\raggedright\arraybackslash}p{0.82\textwidth}@{}}
\toprule
\textbf{Role} & \textbf{Content} \\
\midrule
\rowcolor{AppBand}
User
& \textbf{Query:} Find my posts from Spring Festival about grandpa saying goodnight at Gongheyong with the most comments.\newline
\textbf{Reference time:} Wednesday, March 4, 2026, 04:33. \\
Agent $\cdot$ Step 1
& \emph{``My posts'' maps to self-authorship.}\newline
\texttt{select\_by\_person(relationship\_type="self")} \\
\rowcolor{AppFill}
Observation
& \textbf{8,663 $\rightarrow$ 23.} The self relationship matched. \\
Agent $\cdot$ Step 2
& \emph{Use the canonical Spring Festival annotation rather than a guessed date range.}\newline
\texttt{select\_by\_time(festival="Spring Festival")} \\
\rowcolor{AppFill}
Observation
& \textbf{23 $\rightarrow$ 2.} Spring Festival and its first three day labels were included. \\
Agent $\cdot$ Step 3
& \emph{Apply the Gongheyong POI constraint.}\newline
\texttt{select\_by\_location(keyword="Gongheyong")} \\
\rowcolor{AppFill}
Observation
& \textbf{2 $\rightarrow$ 1.} One Gongheyong-tagged post remained. \\
Agent $\cdot$ Step 4
& \emph{Verify the remembered content about grandpa saying goodnight.}\newline
\texttt{select\_by\_content(semantic\_query="grandpa saying goodnight")} \\
\rowcolor{AppFill}
Observation
& \textbf{1 $\rightarrow$ 1.} The semantic criterion matched. \\
Agent $\cdot$ Step 5
& \emph{Apply the requested comments order.}\newline
\texttt{finalize(sort\_by="comments")} \\
\rowcolor{AppFill}
Final observation
& \textbf{1 $\rightarrow$ 1.} Self-authored; 2026-02-23 02:52:54; text; Gongheyong, Guangdong; ``Grandpa says it's too late to carve wood. Time for bed. Goodnight!''; 0 likes and 1 comment. \\
\rowcolor{AppBand}
Final answer
& Your post from Spring Festival at Gongheyong in Guangdong captures grandpa saying it's too late to carve wood and time for bed, with a goodnight wish [1]. \\
\bottomrule
\end{tabular}
\end{table*}

\begin{table*}[!t]
\caption{Successful multi-result trajectory in which comments ordering is observable after all four selectors retain two eligible posts (task \texttt{c579\_0053}).}
\label{tab:trajectory-c579}
\centering
\footnotesize
\renewcommand{\arraystretch}{1.10}
\setlength{\tabcolsep}{5pt}
\begin{tabular}{@{}>{\raggedright\arraybackslash\sffamily\scriptsize}p{0.13\textwidth}>{\raggedright\arraybackslash}p{0.82\textwidth}@{}}
\toprule
\textbf{Role} & \textbf{Content} \\
\midrule
\rowcolor{AppBand}
User
& \textbf{Query:} posts by ShiRongPo about morning walk to school in Qingdao in the morning with the most comments\newline
\textbf{Reference time:} Wednesday, March 11, 2026, 03:54. \\
Agent $\cdot$ Step 1
& \emph{The query names ShiRongPo.}\newline
\texttt{select\_by\_person(name="ShiRongPo")} \\
\rowcolor{AppFill}
Observation
& \textbf{7,003 $\rightarrow$ 52.} The normalized name matched \texttt{remark\_name=ShiRongPo}. \\
Agent $\cdot$ Step 2
& \emph{The publication-time constraint is morning.}\newline
\texttt{select\_by\_time(time\_of\_day="morning")} \\
\rowcolor{AppFill}
Observation
& \textbf{52 $\rightarrow$ 14.} Morning posts were retained. \\
Agent $\cdot$ Step 3
& \emph{Apply the Qingdao provenance constraint.}\newline
\texttt{select\_by\_location(keyword="Qingdao")} \\
\rowcolor{AppFill}
Observation
& \textbf{14 $\rightarrow$ 5.} Five Qingdao-tagged posts remained. \\
Agent $\cdot$ Step 4
& \emph{Evaluate the remembered morning-walk-to-school subject.}\newline
\texttt{select\_by\_content(semantic\_query="morning walk to school")} \\
\rowcolor{AppFill}
Observation
& \textbf{5 $\rightarrow$ 2.} Two posts matched the semantic criterion. \\
Agent $\cdot$ Step 5
& \emph{Order the two eligible posts by comment count.}\newline
\texttt{finalize(sort\_by="comments")} \\
\rowcolor{AppFill}
Final observation
& \textbf{2 $\rightarrow$ 2.} [1] Video from Laixi No. 1 Middle School, 32 likes and 9 comments; [2] text reflection from the same POI, 22 likes and 4 comments. \\
\rowcolor{AppBand}
Final answer
& Your close college classmate ShiRongPo shared a video of a cold morning ride to school on an electric bike through Laixi, with bare trees and students in winter coats visible on the sidewalk [1]. They also posted a reflective text about walking to school before dawn, imagining utility poles tracing lines in the mist and turning the routine into a quiet expedition [2]. \\
\bottomrule
\end{tabular}
\end{table*}

\begin{table*}[!t]
\caption{Diagnostic trajectory illustrating over-decomposition of a joint content phrase into an additional location constraint (task \texttt{c504\_0015}).}
\label{tab:trajectory-c504}
\centering
\footnotesize
\renewcommand{\arraystretch}{1.10}
\setlength{\tabcolsep}{5pt}
\begin{tabular}{@{}>{\raggedright\arraybackslash\sffamily\scriptsize}p{0.13\textwidth}>{\raggedright\arraybackslash}p{0.82\textwidth}@{}}
\toprule
\textbf{Role} & \textbf{Content} \\
\midrule
\rowcolor{AppBand}
User
& \textbf{Query:} friends posts about recording sunset over Kaiyuan on Saturday evening\newline
\textbf{Reference time:} Wednesday, April 22, 2026, 10:06. \\
Agent $\cdot$ Step 1
& \emph{Restrict the pool to posts from friends.}\newline
\texttt{select\_by\_person(relationship\_type="friend")} \\
\rowcolor{AppFill}
Observation
& \textbf{6,829 $\rightarrow$ 2,444.} Relationship metadata matched. \\
Agent $\cdot$ Step 2
& \emph{Retain posts published on Saturday evenings.}\newline
\texttt{select\_by\_time(weekday="Saturday", time\_of\_day="evening")} \\
\rowcolor{AppFill}
Observation
& \textbf{2,444 $\rightarrow$ 58.} Weekday and evening predicates matched. \\
Agent $\cdot$ Step 3
& \emph{Interpret Kaiyuan as an explicit post-location constraint.}\newline
\texttt{select\_by\_location(keyword="Kaiyuan")} \\
\rowcolor{AppFill}
Observation
& \textbf{58 $\rightarrow$ 1.} One Kaiyuan-tagged post remained. \\
Agent $\cdot$ Step 4
& \emph{Check the remaining post for the sunset-recording subject.}\newline
\texttt{select\_by\_content(semantic\_query="recording sunset")} \\
\rowcolor{AppFill}
Observation
& \textbf{1 $\rightarrow$ 1.} The semantic criterion matched. \\
Agent $\cdot$ Step 5
& \emph{No sort preference is expressed, so use importance.}\newline
\texttt{finalize(sort\_by="importance")} \\
\rowcolor{AppFill}
Final observation
& \textbf{1 $\rightarrow$ 1.} ZhangXiaoLei; 2026-04-18 17:02:57; video; Kaiyuan No. 1 Vocational High School; a phone recording a sunset over Kaiyuan; 0 likes and 0 comments. \\
\rowcolor{AppBand}
Final answer
& Your close friend ZhangXiaoLei shared a video of a smartphone recording a sunset over the streets of Kaiyuan, with a vocational school backpack silhouetted in the foreground while a song plays [1]. \\
\rowcolor{AppStripe}
Diagnostic note
& The gold action sequence omits \texttt{select\_by\_location}: ``recording sunset over Kaiyuan'' is kept as a joint semantic description. The rollout retrieves the target but does not exactly match the intended decomposition. \\
\bottomrule
\end{tabular}
\end{table*}

\begin{table*}[!t]
\caption{Diagnostic trajectory illustrating confusion between a child depicted in a post and the post author's age group (task \texttt{c574\_0006}).}
\label{tab:trajectory-c574}
\centering
\footnotesize
\renewcommand{\arraystretch}{1.10}
\setlength{\tabcolsep}{5pt}
\begin{tabular}{@{}>{\raggedright\arraybackslash\sffamily\scriptsize}p{0.13\textwidth}>{\raggedright\arraybackslash}p{0.82\textwidth}@{}}
\toprule
\textbf{Role} & \textbf{Content} \\
\midrule
\rowcolor{AppBand}
User
& \textbf{Query:} show posts about a child saying goodnight tagged at Kashgar posted late at night with the most likes\newline
\textbf{Reference time:} Thursday, October 23, 2025, 08:38. \\
Agent $\cdot$ Step 1
& \emph{Interpret ``a child'' as an author-age constraint.}\newline
\texttt{select\_by\_person(age\_group="child")} \\
\rowcolor{AppFill}
Observation
& \textbf{5,681 $\rightarrow$ 357.} Child author profiles were retained. \\
Agent $\cdot$ Step 2
& \emph{Map late night to the midnight bucket.}\newline
\texttt{select\_by\_time(time\_of\_day="midnight")} \\
\rowcolor{AppFill}
Observation
& \textbf{357 $\rightarrow$ 56.} Midnight posts were retained. \\
Agent $\cdot$ Step 3
& \emph{Apply the Kashgar location tag.}\newline
\texttt{select\_by\_location(keyword="Kashgar")} \\
\rowcolor{AppFill}
Observation
& \textbf{56 $\rightarrow$ 2.} Two Kashgar-tagged posts remained. \\
Agent $\cdot$ Step 4
& \emph{Evaluate the subject ``child saying goodnight.''}\newline
\texttt{select\_by\_content(semantic\_query="child saying goodnight")} \\
\rowcolor{AppFill}
Observation
& \textbf{2 $\rightarrow$ 1.} The semantic selector retained one post. \\
Agent $\cdot$ Step 5
& \emph{Apply the requested likes order.}\newline
\texttt{finalize(sort\_by="likes")} \\
\rowcolor{AppFill}
Final observation
& \textbf{1 $\rightarrow$ 1.} CaiAiZhi; 2025-08-27 01:56:17; video; Mixia Village, Kashgar; a child waves before Grandma says it is time for bed; 4 likes and 0 comments. \\
\rowcolor{AppBand}
Final answer
& Your distant friend CaiAiZhi shared a late-night video from Mixia Village in Kashgar where a child waves at the camera before running inside as Grandma says it's time for bed and goodnight [1]. \\
\rowcolor{AppStripe}
Diagnostic note
& The gold action sequence begins with time and omits \texttt{select\_by\_person}. The child is the depicted subject, not an explicitly requested author cohort, so the initial age filter is unlicensed. \\
\bottomrule
\end{tabular}
\end{table*}

\begin{table*}[!t]
\caption{System prompt of SocialBuddy, Part I: task definition and decision policy. The instructions define intent decomposition, recursive refinement, parameter grounding, evidence ordering, terminal presentation, rationale generation, and grounded response composition.}
\label{tab:system-prompt-policy}
\ShowcasePanel{SocialBuddy System Prompt: Task and Decision Policy}{%
\ttfamily\raggedright
\setlength{\parindent}{0pt}
\setlength{\parskip}{2pt}
You are SocialBuddy, an autonomous retrieval agent specialized in searching user posts within a social network feed (``Moments''). The user expresses a retrieval objective in natural language. Locate the matching posts and provide a concise, helpful summary.\par

\# Core execution rules\par
1. Intent decomposition. Identify the active constraints across five dimensions: author identity, temporal grounding, spatial provenance, content or modality, and presentation priority.\par
2. Recursive refinement. Compose selection tools with a single functional scope so that each action refines the current candidate set using a criterion supported by the query. Invoke only tools justified by the request: do not introduce unexpressed constraints or omit expressed ones.\par
3. Parameter grounding. Derive all arguments from the query. Resolve relative temporal expressions against the supplied reference timestamp.\par
4. Evidence ordering. Apply \texttt{select\_by\_person}, \texttt{select\_by\_time}, and \texttt{select\_by\_location} before \texttt{select\_by\_content}; conclude every trajectory with exactly one \texttt{finalize} call. Structured metadata constraints precede semantic evaluation over an open vocabulary.\par
5. Terminal presentation. \texttt{finalize} orders the surviving candidates according to \texttt{sort\_by} and returns the top results. Set \texttt{sort\_by} to \texttt{"importance"} unless the query explicitly requests an interaction-based order.\par
6. Rationale and response. Before each action, provide a brief first-person rationale that identifies the supporting query facet and explains the argument derivation. After the terminal \texttt{finalize} action returns its results, produce a concise response that attributes each result to its author and uses exact bracketed citations \texttt{[n]}.}
\end{table*}

\begin{table*}[!t]
\caption{System prompt of SocialBuddy, Part II: citation, serialization of tool calls, operational constraints, and grounding of query context. The placeholder for the tool schema is instantiated by the public action specifications detailed in Tables~\ref{tab:person-parameters}--\ref{tab:remaining-parameters}.}
\label{tab:system-prompt-protocol}
\ShowcasePanel{SocialBuddy System Prompt: Serialization and Grounding Protocol}{%
\ttfamily\raggedright
\setlength{\parindent}{0pt}
\setlength{\parskip}{2pt}
\# Grounding and citation protocol\par
In the final natural language response, cite retrieved posts with square-bracket indices \texttt{[1]}, \texttt{[2]}, \texttt{[3]}, \ldots\ placed at the end of the sentence describing each post. Each bracketed index \texttt{[n]} must sequentially reference the corresponding item in the list returned by \texttt{finalize}. Do not use standard numbered lists as citations.\par

\# Tool schema\par
\textless tools\textgreater\{serialized public action schemas\}\textless/tools\textgreater\par

\# Tool-call formatting interface\par
To invoke a function, emit exactly the following structure. The action message ends with the closing tag:\par
\textless tool\_call\textgreater\par
\textless function=function\_name\textgreater\par
\textless parameter=parameter\_1\textgreater value\_1\textless/parameter\textgreater\par
\textless parameter=parameter\_2\textgreater\par
multi-line parameter value\par
\textless/parameter\textgreater\par
\textless/function\textgreater\par
\textless/tool\_call\textgreater\par

\# Operational constraints\par
- Every function call must conform to the structure above, with a function block nested inside \texttt{tool\_call} tags.\par
- All required arguments must be explicitly populated.\par
- The action rationale must precede the \texttt{tool\_call} block, and each action message must end at its closing tag.\par
- After the environment returns output for the terminal \texttt{finalize} action, provide the natural language summary in the subsequent response.\par
- If no action is required, answer directly from the available context without disclosing the function-calling protocol.\par

\# Context grounding wrapper\par
[Current time] \textless weekday\textgreater, \textless month\textgreater\ \textless day\textgreater, \textless year\textgreater, \textless HH:MM\textgreater\par
[User query] \textless user\_query\textgreater}
\end{table*}

\begin{table*}[!t]
\caption{Structured persona representation for GreenThumbChef, covering demographic context, education, occupation, household structure, daily routine, values, interests, and expression controls.}
\label{tab:persona-showcase}
\centering
\small
\renewcommand{\arraystretch}{1.10}
\setlength{\tabcolsep}{3.2pt}
\begin{tabular}{@{}p{0.15\linewidth}p{0.25\linewidth}p{0.54\linewidth}@{}}
\toprule
\textbf{Facet} & \textbf{Source field(s)} & \textbf{Instantiated value} \\
\midrule
\rowcolor{AppBand}\multicolumn{3}{@{}l}{\textbf{Demographic and setting}}\\
Identity & \texttt{age\_info}, \texttt{gender} & 34-year-old adult woman.\\
\rowcolor{AppStripe}
Location & \texttt{location} & Rural Guizhou, China; the source places her in the Qiandongnan Miao and Dong Autonomous Prefecture.\\
Education & \texttt{education} & Ph.D.; key-university background.\\
\rowcolor{AppBand}\multicolumn{3}{@{}l}{\textbf{Household and work}}\\
Family & \texttt{family} & Married; lives with her spouse, two children, and father-in-law in a rural multigenerational household.\\
\rowcolor{AppStripe}
Occupation & \texttt{career\_info} & Manages human-resources information systems; her narrative context situates this work in a regional agricultural cooperative.\\
Home and mobility & \texttt{living: housing\_type,}\newline\texttt{housing\_description,}\newline\texttt{commute\_method} & Maintained three-story self-built house with a courtyard; e-bike commute.\\
\rowcolor{AppStripe}
Daily routine & \texttt{living.daily\_routine} & Prepares breakfast, takes the children to school, manages HR data, and returns to cooking and elder care.\\
\rowcolor{AppBand}\multicolumn{3}{@{}l}{\textbf{Disposition}}\\
Values & \texttt{personal\_values} & Pragmatic balance between career efficiency and family stability.\\
\rowcolor{AppStripe}
Attitude & \texttt{life\_attitude: attitude,}\newline\texttt{attitude\_details} & Grounded optimism: careful scheduling and steady progress reconcile professional and household demands.\\
Coping & \texttt{life\_attitude.}\newline\texttt{coping\_mechanism} & Prioritizes tasks logically and uses brief quiet moments to reset.\\
\rowcolor{AppBand}\multicolumn{3}{@{}l}{\textbf{Interests and social expression}}\\
Interests & \texttt{interests} & Gardening and cooking.\\
\rowcolor{AppStripe}
Public identity & \texttt{posting\_style:}\newline\texttt{screen\_name} & GreenThumbChef.\\
Expression controls & \texttt{posting\_style} & Life Logger; calm and restrained tone; several posts per week.\\
\rowcolor{AppStripe}
Topic priors & \texttt{common\_topics}, \texttt{style\_description} & Home gardening, home cooking, work--life balance, family moments, and seasonal activities; grounded updates connect career demands with family stability.\\
\rowcolor{AppBand}\multicolumn{3}{@{}l}{\textbf{Narrative and network grounding}}\\
Narrative anchors & \texttt{personal\_story} & HRIS updates, school runs, courtyard maintenance, cooking, elder care, a power outage, and methodical problem solving.\\
\rowcolor{AppStripe}
Ego relation & \texttt{social\_info:}\newline\texttt{relationship\_type, closeness} & Neighbor; close tie.\\
\bottomrule
\end{tabular}
\end{table*}

\begin{table*}[!t]
\caption{Structured persona representation for Grandma Li, characterized by late-life community participation, intergenerational care, and reflective expression.}
\label{tab:persona-grandma-li}
\centering
\small
\renewcommand{\arraystretch}{1.10}
\setlength{\tabcolsep}{3.2pt}
\begin{tabular}{@{}p{0.15\linewidth}p{0.25\linewidth}p{0.54\linewidth}@{}}
\toprule
\textbf{Facet} & \textbf{Source field(s)} & \textbf{Instantiated value} \\
\midrule
\rowcolor{AppBand}\multicolumn{3}{@{}l}{\textbf{Demographic and setting}}\\
Identity & \texttt{age\_info}, \texttt{gender} & 81-year-old senior woman.\\
\rowcolor{AppStripe}
Location & \texttt{location} & Chongqing, China.\\
Education & \texttt{education} & Bachelor's education; public-school background.\\
\rowcolor{AppBand}\multicolumn{3}{@{}l}{\textbf{Household and work}}\\
Family & \texttt{family} & Widowed; two children; lives with her son's family, including an adult grandson, in a multigenerational household.\\
\rowcolor{AppStripe}
Occupation & \texttt{career\_info} & Retired community volunteer.\\
Home and mobility & \texttt{living: housing\_type,}\newline\texttt{housing\_description,}\newline\texttt{commute\_method} & Lives with family in a three-bedroom apartment in an older building; travels locally on foot.\\
\rowcolor{AppStripe}
Daily routine & \texttt{living.daily\_routine} & Practices early-morning Tai Chi, helps prepare meals, and participates in household routines.\\
\rowcolor{AppBand}\multicolumn{3}{@{}l}{\textbf{Disposition}}\\
Values & \texttt{personal\_values} & Community service, family unity, compassionate stewardship, resilience, and hope.\\
\rowcolor{AppStripe}
Attitude & \texttt{life\_attitude: attitude,}\newline\texttt{attitude\_details} & Hopeful and grateful; understands daily kindness and intergenerational care as continuations of lifelong service.\\
Coping & \texttt{life\_attitude.}\newline\texttt{coping\_mechanism} & Draws strength from stable routines and the presence of her son's family.\\
\rowcolor{AppBand}\multicolumn{3}{@{}l}{\textbf{Interests and social expression}}\\
Interests & \texttt{interests} & Tai Chi and traditional Chinese cooking.\\
\rowcolor{AppStripe}
Public identity & \texttt{posting\_style:}\newline\texttt{screen\_name} & Grandma Li.\\
Expression controls & \texttt{posting\_style} & Family Elder; warm and gentle tone; several posts per week.\\
\rowcolor{AppStripe}
Topic priors & \texttt{common\_topics}, \texttt{style\_description} & Family gatherings, traditional cooking, health tips, festival greetings, and Tai Chi; a gentle, grateful family/community voice.\\
\rowcolor{AppBand}\multicolumn{3}{@{}l}{\textbf{Narrative and network grounding}}\\
Narrative anchors & \texttt{personal\_story} & Park Tai Chi, cooking, charity coordination, neighborhood mediation, dumpling making, and community-event logistics.\\
\rowcolor{AppStripe}
Ego relation & \texttt{social\_info:}\newline\texttt{relationship\_type, closeness} & Friend; normal tie.\\
\bottomrule
\end{tabular}
\end{table*}

\begin{table*}[!t]
\caption{Structured persona representation for Steady Hearth, coupling multigenerational family routines with responsible media supervision.}
\label{tab:persona-steady-hearth}
\centering
\small
\renewcommand{\arraystretch}{1.10}
\setlength{\tabcolsep}{3.2pt}
\begin{tabular}{@{}p{0.15\linewidth}p{0.25\linewidth}p{0.54\linewidth}@{}}
\toprule
\textbf{Facet} & \textbf{Source field(s)} & \textbf{Instantiated value} \\
\midrule
\rowcolor{AppBand}\multicolumn{3}{@{}l}{\textbf{Demographic and setting}}\\
Identity & \texttt{age\_info}, \texttt{gender} & 54-year-old middle-aged woman.\\
\rowcolor{AppStripe}
Location & \texttt{location} & Weifang, Shandong, China.\\
Education & \texttt{education} & College education; public-school background.\\
\rowcolor{AppBand}\multicolumn{3}{@{}l}{\textbf{Household and work}}\\
Family & \texttt{family} & Married; lives with her spouse, adult son, and young granddaughter in a multigenerational household.\\
\rowcolor{AppStripe}
Occupation & \texttt{career\_info} & Media manager/supervisor.\\
Home and mobility & \texttt{living: housing\_type,}\newline\texttt{housing\_description,}\newline\texttt{commute\_method} & Owns a three-bedroom apartment in a modern mid-rise building; e-bike commute.\\
\rowcolor{AppStripe}
Daily routine & \texttt{living.daily\_routine} & Prepares breakfast, takes her granddaughter to kindergarten, supervises media work, shops locally, and cooks for the family.\\
\rowcolor{AppBand}\multicolumn{3}{@{}l}{\textbf{Disposition}}\\
Values & \texttt{personal\_values} & Family stability, media integrity, and responsible leadership.\\
\rowcolor{AppStripe}
Attitude & \texttt{life\_attitude: attitude,}\newline\texttt{attitude\_details} & Grounded optimism; steady effort in supervisory work and family care yields meaningful stability.\\
Coping & \texttt{life\_attitude.}\newline\texttt{coping\_mechanism} & Relies on open communication and multigenerational support.\\
\rowcolor{AppBand}\multicolumn{3}{@{}l}{\textbf{Interests and social expression}}\\
Interests & \texttt{interests} & Cooking traditional meals and relaxing on the balcony in the evening.\\
\rowcolor{AppStripe}
Public identity & \texttt{posting\_style:}\newline\texttt{screen\_name} & Steady Hearth.\\
Expression controls & \texttt{posting\_style} & Family Anchor; warm and gentle tone; weekly posting.\\
\rowcolor{AppStripe}
Topic priors & \texttt{common\_topics}, \texttt{style\_description} & Home cooking, family gatherings, evening reflections, responsible media, and seasonal greetings; warm domestic observations remain tied to integrity.\\
\rowcolor{AppBand}\multicolumn{3}{@{}l}{\textbf{Narrative and network grounding}}\\
Narrative anchors & \texttt{personal\_story} & Reviewing agricultural promotional videos for factual integrity, mentoring younger editors, handling a server outage, and supporting household transitions.\\
\rowcolor{AppStripe}
Ego relation & \texttt{social\_info:}\newline\texttt{relationship\_type, closeness} & Business partner; close tie.\\
\bottomrule
\end{tabular}
\end{table*}

\begin{table*}[!t]
\caption{Structured persona representation for Harmony Home, connecting ethical healthcare communication with multigenerational care.}
\label{tab:persona-harmony-home}
\centering
\small
\renewcommand{\arraystretch}{1.10}
\setlength{\tabcolsep}{3.2pt}
\begin{tabular}{@{}p{0.15\linewidth}p{0.25\linewidth}p{0.54\linewidth}@{}}
\toprule
\textbf{Facet} & \textbf{Source field(s)} & \textbf{Instantiated value} \\
\midrule
\rowcolor{AppBand}\multicolumn{3}{@{}l}{\textbf{Demographic and setting}}\\
Identity & \texttt{age\_info}, \texttt{gender} & 51-year-old middle-aged woman.\\
\rowcolor{AppStripe}
Location & \texttt{location} & Chongqing, China.\\
Education & \texttt{education} & Bachelor's education; private-school background.\\
\rowcolor{AppBand}\multicolumn{3}{@{}l}{\textbf{Household and work}}\\
Family & \texttt{family} & Married; lives with her spouse, adult daughter, and elderly mother in a multigenerational urban household.\\
\rowcolor{AppStripe}
Occupation & \texttt{career\_info} & Pharmaceutical marketing specialist/assistant.\\
Home and mobility & \texttt{living: housing\_type,}\newline\texttt{housing\_description,}\newline\texttt{commute\_method} & Three-bedroom apartment in an older building; bus commute.\\
\rowcolor{AppStripe}
Daily routine & \texttt{living.daily\_routine} & Prepares breakfast for her mother and daughter, travels to administrative work by bus, and returns to prepare dinner.\\
\rowcolor{AppBand}\multicolumn{3}{@{}l}{\textbf{Disposition}}\\
Values & \texttt{personal\_values} & Ethical healthcare advancement, family stability, and community well-being.\\
\rowcolor{AppStripe}
Attitude & \texttt{life\_attitude: attitude,}\newline\texttt{attitude\_details} & Hopeful and pragmatic; balances meaningful healthcare communication with care for her daughter and mother.\\
Coping & \texttt{life\_attitude.}\newline\texttt{coping\_mechanism} & Uses open communication with her spouse and draws strength from incremental progress at work and home.\\
\rowcolor{AppBand}\multicolumn{3}{@{}l}{\textbf{Interests and social expression}}\\
Interests & \texttt{interests} & Cooking and watching historical dramas.\\
\rowcolor{AppStripe}
Public identity & \texttt{posting\_style:}\newline\texttt{screen\_name} & Harmony Home.\\
Expression controls & \texttt{posting\_style} & Family Guardian; warm and gentle tone; weekly posting.\\
\rowcolor{AppStripe}
Topic priors & \texttt{common\_topics}, \texttt{style\_description} & Home cooking, family gatherings, health and wellness, historical dramas, and community events; warm family observations emphasize ethics and stability.\\
\rowcolor{AppBand}\multicolumn{3}{@{}l}{\textbf{Narrative and network grounding}}\\
Narrative anchors & \texttt{personal\_story} & Compliance-oriented medical communication, clearer patient information, an adult child's career transition, elder care, and public-transit routines.\\
\rowcolor{AppStripe}
Ego relation & \texttt{social\_info:}\newline\texttt{relationship\_type, closeness} & Friend; normal tie.\\
\bottomrule
\end{tabular}
\end{table*}

\begin{table*}[!t]
\caption{Structured persona representation for Practical Bloom, organized around urban work logistics, child care, gardening, and cooking.}
\label{tab:persona-practical-bloom}
\centering
\small
\renewcommand{\arraystretch}{1.10}
\setlength{\tabcolsep}{3.2pt}
\begin{tabular}{@{}p{0.15\linewidth}p{0.25\linewidth}p{0.54\linewidth}@{}}
\toprule
\textbf{Facet} & \textbf{Source field(s)} & \textbf{Instantiated value} \\
\midrule
\rowcolor{AppBand}\multicolumn{3}{@{}l}{\textbf{Demographic and setting}}\\
Identity & \texttt{age\_info}, \texttt{gender} & 32-year-old adult woman.\\
\rowcolor{AppStripe}
Location & \texttt{location} & Coastal Guangdong, China.\\
Education & \texttt{education} & College education; private-school background.\\
\rowcolor{AppBand}\multicolumn{3}{@{}l}{\textbf{Household and work}}\\
Family & \texttt{family} & Married; lives with her spouse, young daughter, and mother-in-law in a multigenerational urban household.\\
\rowcolor{AppStripe}
Occupation & \texttt{career\_info} & Parts sales.\\
Home and mobility & \texttt{living: housing\_type,}\newline\texttt{housing\_description,}\newline\texttt{commute\_method} & Owns a well-lit apartment in a modern residential complex; e-bike commute.\\
\rowcolor{AppStripe}
Daily routine & \texttt{living.daily\_routine} & Takes her daughter to kindergarten, works at a local office, picks her up, and cooks with her mother-in-law.\\
\rowcolor{AppBand}\multicolumn{3}{@{}l}{\textbf{Disposition}}\\
Values & \texttt{personal\_values} & Family stability and practical daily solutions.\\
\rowcolor{AppStripe}
Attitude & \texttt{life\_attitude: attitude,}\newline\texttt{attitude\_details} & Pragmatic optimism: steady effort, family unity, and efficient household logistics support a stable life.\\
Coping & \texttt{life\_attitude.}\newline\texttt{coping\_mechanism} & Uses open communication and decomposes problems into manageable steps.\\
\rowcolor{AppBand}\multicolumn{3}{@{}l}{\textbf{Interests and social expression}}\\
Interests & \texttt{interests} & Cooking and gardening.\\
\rowcolor{AppStripe}
Public identity & \texttt{posting\_style:}\newline\texttt{screen\_name} & Practical Bloom.\\
Expression controls & \texttt{posting\_style} & Life Logger; warm and gentle tone; several posts per week.\\
\rowcolor{AppStripe}
Topic priors & \texttt{common\_topics}, \texttt{style\_description} & Home cooking, gardening tips, family moments, daily routines, and seasonal harvests; a grounded domestic voice reflects pragmatic optimism.\\
\rowcolor{AppBand}\multicolumn{3}{@{}l}{\textbf{Narrative and network grounding}}\\
Narrative anchors & \texttt{personal\_story} & Coastal commuting, seafood shopping, coordinating parts shipments for fishing and agricultural equipment, delivery rescheduling, and family dinner.\\
\rowcolor{AppStripe}
Ego relation & \texttt{social\_info:}\newline\texttt{relationship\_type, closeness} & Colleague; close tie.\\
\bottomrule
\end{tabular}
\end{table*}

\begin{table*}[!t]
\caption{Field composition and visibility boundary of a SocialEnv persona record. Public author attributes support person-level retrieval, whereas private context conditions post synthesis but is not directly exposed to the search agent.}
\label{tab:persona-record-composition}
\centering
\footnotesize
\renewcommand{\arraystretch}{1.08}
\setlength{\tabcolsep}{4pt}
\begin{tabular}{@{}p{0.18\linewidth}p{0.38\linewidth}p{0.17\linewidth}p{0.20\linewidth}@{}}
\toprule
\textbf{Field group} & \textbf{Included attributes} & \textbf{Role in synthesis} & \textbf{Exposure to social search} \\
\midrule
\rowcolor{AppBand}\multicolumn{4}{@{}l}{\textbf{Public author layer}}\\
Author attributes
& Age or age group, gender, surface identity, relationship type, and closeness.
& Conditions author identity and voice.
& Authorized categorical subset consumed by \texttt{select\_by\_person}. \\
\rowcolor{AppBand}\multicolumn{4}{@{}l}{\textbf{Private generation context}}\\
\rowcolor{AppStripe}
Background context
& Education, occupation, coarse home region, household composition, housing, and mobility.
& Establishes persistent lived context.
& Not directly exposed. \\
Behavioral context
& Daily routine, values, life attitude, coping mechanism, interests, and narrative anchors.
& Governs recurring activities, concerns, and interpretations.
& Not directly exposed. \\
\rowcolor{AppStripe}
Expression controls
& Screen name, posting archetype, tone, frequency, common-topic priors, and style description.
& Regulates topic sampling and linguistic realization.
& Not directly exposed as selector predicates. \\
\rowcolor{AppBand}\multicolumn{4}{@{}l}{\textbf{Operational linkage}}\\
Author--history association
& Internal profile key, author--post association, and ego-circle membership.
& Maintains authorship and visibility consistency across modalities.
& Pipeline state only; not semantic evidence. \\
\bottomrule
\end{tabular}
\end{table*}

\onecolumn
\begingroup
\footnotesize
\renewcommand{\arraystretch}{1.08}
\setlength{\tabcolsep}{3.4pt}
\setlength{\LTleft}{0pt}
\setlength{\LTright}{0pt}
\begin{longtable}{@{}p{0.09\linewidth}p{0.20\linewidth}p{0.46\linewidth}p{0.19\linewidth}@{}}
\caption{Cross-modal post records for Persona A (GreenThumbChef). ``Payload'' reports each applicable modality-specific field; \emph{none} marks nullable fields, and long media descriptions or supplements are summarized.}
\label{tab:post-showcase}\\
\toprule
\textbf{Type / ID} & \textbf{Time, context, engagement} & \textbf{Authored text and payload} & \textbf{Persona realization} \\
\midrule
\endfirsthead
\endhead
\endfoot
\bottomrule
\endlastfoot
\rowcolor{AppBand}\multicolumn{4}{@{}l}{\textbf{A1 --- Gardening milestone}}\\*
\texttt{text}\newline 005
& 2025-08-04 15:48:47\newline Monday; festival: none\newline Topic: gardening\newline POI: none\newline 30 likes; 7 comments
& \textbf{Post:} ``The first tomato finally ripened on the vine this afternoon. A small, red milestone after months of tending.''\newline
\textbf{Payload:} Media/share: none.
& Gardening is explicit; the measured ``milestone'' and restrained two-sentence caption instantiate patient maintenance and steady progress. \\
\rowcolor{AppBand}\multicolumn{4}{@{}l}{\textbf{A2 --- Rural commute pause}}\\*
\rowcolor{AppStripe}
\texttt{image}\newline 012
& 2025-08-27 17:23:00\newline Wednesday; festival: none\newline Topic: cloud notes\newline POI city: Long'e County\newline POI name: withheld\newline 40 likes; 5 comments
& \textbf{Post:} ``5:23 PM. The clouds over Long'e are putting on a show again. Just stopped the e-bike to look up before heading home to the kids.''\newline
\textbf{Payload:} 9 images; video count: none. Descriptions cover sunset clouds, an e-bike on a rural road, weather display, karst mountains, a wedding band, power lines, a puddle reflection, cloud detail, and fields at the horizon.
& The rural commute, e-bike, return to her children, and deliberate pause jointly expose daily routine and the stated reflective coping pattern. \\
\rowcolor{AppBand}\multicolumn{4}{@{}l}{\textbf{A3 --- Dawn household routine}}\\*
\texttt{video}\newline 010
& 2025-08-16 05:30:55\newline Saturday; festival: none\newline Topic: home cooking\newline POI: none\newline 12 likes; 0 comments
& \textbf{Post:} ``5:30 AM. The kitchen is quiet while everyone sleeps. Starting the congee before the morning rush begins.''\newline
\textbf{Payload:} Image count: none; 1 video describing rice congee simmering on a gas stove in a dim kitchen, with steam rising from the pot.
& Early meal preparation materializes planning and care for a multigenerational household; the precise timestamp and wording remain methodical. \\
\rowcolor{AppBand}\multicolumn{4}{@{}l}{\textbf{A4 --- Work--family music share}}\\*
\rowcolor{AppStripe}
\texttt{share\_}\newline\texttt{music}\newline 034
& 2025-12-22 22:33:35\newline Monday; festival: none\newline Topic: music\newline POI: none\newline 33 likes; 3 comments
& \textbf{Post:} ``Finally, the house is quiet. After managing HR data all day and tucking the kids in, Zhao Zhao's `Ru Yan' hits different. Proud of this balanced life I've built in Long'e.''\newline
\textbf{Payload:} \textit{Ru Yan}, artist Zhao Zhao; the supplement identifies it as a song performed by Zhao Zhao. Media: none.
& The record directly joins HRIS work, child care, work--family balance, and a quiet evening reset. \\
\rowcolor{AppBand}\multicolumn{4}{@{}l}{\textbf{A5 --- Agricultural technology reading}}\\*
\texttt{share\_}\newline\texttt{article}\newline 024
& 2025-11-09 08:36:47\newline Sunday; festival: none\newline Topic: reading\newline POI: none\newline 19 likes; 5 comments
& \textbf{Post:} ``Sunday morning read! Even in HRIS, I admire the tech behind these agricultural drones. Smart farming is the future!''\newline
\textbf{Payload:} \textit{Product Recommendation: Full-Constellation, Full-Frequency Helical Antenna for Crop-Protection Drones} (translated title); preview concerns stable reception around trees and in humid or dusty fields. Artist/media: none.
& Information-systems expertise transfers to an agricultural technology problem grounded in the persona's rural context. \\
\rowcolor{AppBand}\multicolumn{4}{@{}l}{\textbf{A6 --- Bedtime study check}}\\*
\texttt{video}\newline 003
& 2025-07-20 23:44:30\newline Sunday; festival: none\newline Topic: study\newline POI: none\newline 11 likes; 1 comment
& \textbf{Post:} ``Finally quiet. Watching the kids' study notes one last time before bed. May their dreams be as bright as their potential.''\newline
\textbf{Payload:} Image count: none; 1 video describing a warm desk lamp over open notebooks and children's drawings as a hand turns a page.
& The bedtime review makes educational care and the late-evening household routine observable in a restrained register. \\
\rowcolor{AppBand}\multicolumn{4}{@{}l}{\textbf{A7 --- Wetland lunch-break walk}}\\*
\rowcolor{AppStripe}
\texttt{image}\newline 007
& 2025-08-05 14:29:41\newline Tuesday; festival: none\newline Topic: wetlands\newline POI: none\newline 25 likes; 4 comments
& \textbf{Post:} ``Lunch break stroll near the shop. The August heat makes the wetland reeds grow wild and green.''\newline
\textbf{Payload:} 9 images; video count: none. Descriptions cover reeds, a boardwalk, water reflections, dragonfly and flower details, fish shadows, lotus leaves, and the town skyline.
& A workday pause in a local landscape extends the rural setting and the persona's preference for brief restorative observation. \\
\rowcolor{AppBand}\multicolumn{4}{@{}l}{\textbf{A8 --- Evening home maintenance}}\\*
\texttt{text}\newline 014
& 2025-09-19 19:20:39\newline Friday; festival: none\newline Topic: curtains\newline POI city: Long'e\newline POI name: residential area (coarsened)\newline 28 likes; 5 comments
& \textbf{Post:} ``Finally hung the new curtains after the kids fell asleep. Quiet evening at last.''\newline
\textbf{Payload:} Media/share: none.
& Home maintenance after child care exemplifies practical household management and the recurrent quiet-evening cadence. \\
\rowcolor{AppBand}\multicolumn{4}{@{}l}{\textbf{A9 --- Balcony reading corner}}\\*
\rowcolor{AppStripe}
\texttt{text}\newline 015
& 2025-09-28 23:36:05\newline Sunday; festival: none\newline Topic: balcony makeover\newline POI: none\newline 28 likes; 3 comments
& \textbf{Post:} ``Finally finished our balcony makeover! A perfect little spot for my PhD books and evening tea after the kids are asleep. So proud of this cozy corner!''\newline
\textbf{Payload:} Media/share: none.
& The renovated corner joins advanced education, domestic space, child care, and a deliberate evening reset. \\
\rowcolor{AppBand}\multicolumn{4}{@{}l}{\textbf{A10 --- Work--family pause in autumn}}\\*
\texttt{video}\newline 019
& 2025-10-14 14:48:42\newline Tuesday; festival: none\newline Topic: love diary\newline POI: none\newline 12 likes; 2 comments
& \textbf{Post:} ``Between spreadsheets and school runs, I found a pause. The ginkgo leaves are turning gold, whispering that time, like love, is best measured in quiet moments.''\newline
\textbf{Payload:} Image count: none; 1 video describing golden ginkgo leaves moving against a blue afternoon sky.
& Spreadsheets and school runs establish the dual role, while the visual pause instantiates the stated reflective coping pattern. \\
\rowcolor{AppBand}\multicolumn{4}{@{}l}{\textbf{A11 --- Family cooking humor}}\\*
\texttt{text}\newline 020
& 2025-10-16 10:19:10\newline Thursday; festival: none\newline Topic: home cooking\newline POI: none\newline 9 likes; 0 comments
& \textbf{Post:} ``My 5yo declared my braised pork `too chewy for her royal teeth.' Guess I'm demoted from Head Chef to Snack Vendor.''\newline
\textbf{Payload:} Media/share: none.
& Cooking and the five-year-old child's perspective appear through mild domestic humor rather than an explicit persona restatement. \\
\rowcolor{AppBand}\multicolumn{4}{@{}l}{\textbf{A12 --- HRIS workstation}}\\*
\rowcolor{AppStripe}
\texttt{image}\newline 036
& 2025-12-30 10:20:23\newline Tuesday; festival: none\newline Topic: music\newline POI city: Liupanshui\newline POI name: HR workplace (coarsened)\newline 17 likes; 2 comments
& \textbf{Post:} ``Tuesday morning hum. Coffee in hand, systems running.''\newline
\textbf{Payload:} 9 images; video count: none. Descriptions show HRIS code and dashboards, an employee-data binder, notebook and files, coffee, headphones, a small plant, and an office-window view.
& Technical workplace evidence grounds the HRIS occupation, while the compact caption preserves routine-oriented narration. \\
\rowcolor{AppBand}\multicolumn{4}{@{}l}{\textbf{A13 --- New Year garden greeting}}\\*
\texttt{video}\newline 045
& 2026-02-17 07:31:41\newline Tuesday; Spring Festival (1st Day)\newline Topic: gardening\newline POI: none\newline 27 likes; 2 comments
& \textbf{Post:} ``Happy Lunar New Year! Wishing our family health and joy as we start the Year of the Horse.''\newline
\textbf{Payload:} Image count: none; 1 video describing red plum blossoms in a garden under soft morning light.
& Garden imagery, a seasonal greeting, and family-centered wishes jointly instantiate declared interests and topic priors. \\
\rowcolor{AppBand}\multicolumn{4}{@{}l}{\textbf{A14 --- Parenting through steady routines}}\\*
\rowcolor{AppStripe}
\texttt{image}\newline 046
& 2026-02-23 09:47:32\newline Monday; Spring Festival\newline Topic: parenting\newline POI city: Long'e\newline POI name: school (coarsened)\newline 20 likes; 2 comments
& \textbf{Post:} ``The morning rush is finally over. Dropping the kids off on the e-bike, I realized parenting isn't about perfection, but presence. Watching them run into school, I see the quiet result of years of small, consistent moments. A PhD taught me theory; two children taught me life.''\newline
\textbf{Payload:} 9 images; video count: none. Descriptions cover an e-bike and helmets, breakfast and lunchboxes, two children at school, an HRIS calendar alert, elder care, and a misty local street.
& Education, two children, e-bike commuting, meal preparation, elder care, and HRIS scheduling converge in a single source record. \\
\rowcolor{AppBand}\multicolumn{4}{@{}l}{\textbf{A15 --- Family breakfast}}\\*
\texttt{video}\newline 049
& 2026-03-09 07:27:34\newline Monday; festival: none\newline Topic: family recipes\newline POI: none\newline 16 likes; 0 comments
& \textbf{Post:} ``Steam rising, heart full. This taste is home.''\newline
\textbf{Payload:} Image count: none; 1 video describing hot soy milk and fried dough sticks on a wooden table as a child's hands reach for a bowl.
& The concise breakfast scene closes the history with home cooking, child care, and the understated affect of the Life Logger style. \\
\end{longtable}
\endgroup

\begingroup
\footnotesize
\renewcommand{\arraystretch}{1.08}
\setlength{\tabcolsep}{3.4pt}
\setlength{\LTleft}{0pt}
\setlength{\LTright}{0pt}
\begin{longtable}{@{}p{0.09\linewidth}p{0.20\linewidth}p{0.46\linewidth}p{0.19\linewidth}@{}}
\caption{Cross-modal post records for Persona B (Grandma Li).}
\label{tab:post-showcase-grandma-li}\\
\toprule
\textbf{Type / ID} & \textbf{Time, context, engagement} & \textbf{Authored text and payload} & \textbf{Persona realization} \\
\midrule
\endfirsthead
\endhead
\endfoot
\bottomrule
\endlastfoot
\rowcolor{AppBand}\multicolumn{4}{@{}l}{\textbf{B1 --- Traditional family lunch}}\\*
\texttt{text}\newline 012
& 2025-09-29 12:21:45\newline Monday; festival: none\newline Topic: Cooking traditional Chinese cuisine\newline POI: none\newline 44 likes; 2 comments
& \textbf{Post:} ``Lunch is ready: braised pork and steamed fish. The grandson loved it. A quiet Monday at home.''\newline
\textbf{Payload:} Media/share: none.
& Traditional cooking and care for her adult grandson appear in a concise domestic update consistent with a gentle Family Elder voice. \\
\rowcolor{AppBand}\multicolumn{4}{@{}l}{\textbf{B2 --- Reciprocal digital learning}}\\*
\rowcolor{AppStripe}
\texttt{image}\newline 042
& 2026-05-08 17:12:36\newline Friday; festival: none\newline Topic: iPad\newline POI: none\newline 26 likes; 7 comments
& \textbf{Post:} ``Late afternoon light filters through the window, warming the screen where my grandson teaches me to paint. Time slows down in these quiet, colorful moments.''\newline
\textbf{Payload:} 9 images; video count: none. Descriptions show a tablet, stylus and watercolor interface, an elderly learner and younger guide, plum-blossom painting, tea, warm window light, a sketch gallery, and the pair seated together.
& Reciprocal learning reverses the usual direction of care while preserving intergenerational continuity and openness to new practices. \\
\rowcolor{AppBand}\multicolumn{4}{@{}l}{\textbf{B3 --- Tai Chi routine}}\\*
\texttt{video}\newline 002
& 2025-07-04 19:08:25\newline Friday; festival: none\newline Topic: Tai Chi\newline POI: none\newline 30 likes; 8 comments
& \textbf{Post:} ``Even at 81, my Tai Chi flows strong in the summer evening breeze! So proud of keeping active for my family.''\newline
\textbf{Payload:} Image count: none; 1 video describing an elderly woman performing slow Tai Chi movements in a community park at dusk.
& Age, established exercise routine, community setting, and family-oriented motivation are simultaneously observable. \\
\rowcolor{AppBand}\multicolumn{4}{@{}l}{\textbf{B4 --- Memory through music}}\\*
\rowcolor{AppStripe}
\texttt{share\_}\newline\texttt{music}\newline 004
& 2025-07-25 07:55:17\newline Friday; festival: none\newline Topic: music\newline POI city: Chongqing\newline POI name: park (coarsened)\newline 39 likes; 6 comments
& \textbf{Post:} ``Early morning tai chi done. This instrumental reminds me of dancing with my late husband in the park decades ago. The melody is simple, just like our life was. Missing him as the sun rises over Chongqing.''\newline
\textbf{Payload:} \textit{Yeren (Instrumental)} (transliterated title), artists Tang Hanxiao and Zhang Jiaming; the supplement identifies the performers and associated album. Media: none.
& The share links morning routine and widowhood to a gentle coping narrative grounded in long family memory. \\
\rowcolor{AppBand}\multicolumn{4}{@{}l}{\textbf{B5 --- Intergenerational guidance}}\\*
\texttt{share\_}\newline\texttt{article}\newline 017
& 2025-10-26 22:26:43\newline Sunday; festival: none\newline Topic: reading\newline POI: none\newline 41 likes; 3 comments
& \textbf{Post:} ``Late night reading brought back memories. In 2020, even small corrections in the college guide mattered so much to families. Now my grandson is grown, but the lesson remains: clarity comes with time and careful review. Good night.''\newline
\textbf{Payload:} \textit{Important Notice from the Shanxi Provincial Admissions and Examination Center: Corrections to the College Application Guide} (translated title); preview summarizes corrections requested for the 2020 guide. Artist/media: none.
& Administrative correction becomes a reflection on careful stewardship and the grandson's transition to adulthood. \\
\rowcolor{AppBand}\multicolumn{4}{@{}l}{\textbf{B6 --- Riverside family memory}}\\*
\texttt{text}\newline 003
& 2025-07-22 20:12:07\newline Tuesday; festival: none\newline Topic: paddleboarding\newline POI city: Chongqing\newline POI name: riverside park (coarsened)\newline 29 likes; 5 comments
& \textbf{Post:} ``Watching the river flow tonight, I remembered my late husband teaching our son to swim here decades ago. The water is the same, but the hands that held him are gone. A gentle ache in this summer breeze.''\newline
\textbf{Payload:} Media/share: none.
& Place anchors a three-generation memory, making widowhood and intergenerational continuity visible through a measured reflection. \\
\rowcolor{AppBand}\multicolumn{4}{@{}l}{\textbf{B7 --- Dawn community routine}}\\*
\rowcolor{AppStripe}
\texttt{video}\newline 005
& 2025-08-02 04:23:33\newline Saturday; festival: none\newline Topic: Tai Chi\newline POI: none\newline 12 likes; 0 comments
& \textbf{Post:} ``Early morning Tai Chi in the park. The air is fresh and cool. Wishing my family and neighbors a peaceful Saturday filled with health and joy.''\newline
\textbf{Payload:} Image count: none; 1 video describing an elderly woman performing slow Tai Chi movements in a park at dawn.
& The recurring exercise routine expands from family care to neighborhood well-being, consistent with a retired community volunteer. \\
\rowcolor{AppBand}\multicolumn{4}{@{}l}{\textbf{B8 --- Grandson's educational milestone}}\\*
\texttt{text}\newline 009
& 2025-09-11 21:02:16\newline Thursday; festival: none\newline Topic: growth retrospective\newline POI city: Chongqing\newline POI name: withheld\newline 24 likes; 5 comments
& \textbf{Post:} ``My grandson just got his master's degree! At 81, seeing him succeed is my greatest joy. So proud of our family!''\newline
\textbf{Payload:} Media/share: none.
& Her age, the adult grandson, and explicit family pride place educational attainment within an intergenerational life course. \\
\rowcolor{AppBand}\multicolumn{4}{@{}l}{\textbf{B9 --- Pipa at family dinner}}\\*
\rowcolor{AppStripe}
\texttt{video}\newline 018
& 2025-11-08 18:48:58\newline Saturday; festival: none\newline Topic: pipa\newline POI city: Chongqing\newline POI name: omitted\newline 32 likes; 7 comments
& \textbf{Post:} ``On this quiet Saturday evening, my grandson played the pipa for our family dinner. A milestone moment to cherish.''\newline
\textbf{Payload:} Image count: none; 1 video describing a 24-year-old man playing pipa in a living room while an elderly woman and a middle-aged couple watch.
& The visual composition grounds the three household generations, and traditional music becomes shared family practice. \\
\rowcolor{AppBand}\multicolumn{4}{@{}l}{\textbf{B10 --- Reciprocal Winter Solstice care}}\\*
\texttt{image}\newline 024
& 2025-12-22 21:43:53\newline Monday; festival: none\newline Topic: seafood\newline POI: none\newline 20 likes; 1 comment
& \textbf{Post:} ``Winter Solstice dinner with the family. My son and daughter-in-law prepared fresh seafood to celebrate the longest night. Grateful for this warm gathering.''\newline
\textbf{Payload:} 1 image; video count: none. The description shows steaming shrimp and crab on a family dining table.
& Care is reciprocal rather than one-directional: the younger generation prepares the meal, while gratitude sustains the Family Elder voice. \\
\rowcolor{AppBand}\multicolumn{4}{@{}l}{\textbf{B11 --- Slow-and-steady breakfast humor}}\\*
\texttt{text}\newline 025
& 2025-12-29 05:34:39\newline Monday; festival: none\newline Topic: New Year\newline POI city: Chongqing\newline POI name: park (coarsened)\newline 34 likes; 5 comments
& \textbf{Post:} ``Tai Chi at 5:30 AM! My grandson says I move slower than a sleepy snail. I told him, `Slow and steady wins the race to the hot breakfast!''' \newline
\textbf{Payload:} Media/share: none.
& Gentle teasing connects the early exercise routine, shared breakfast, and a relaxed relationship with her adult grandson. \\
\rowcolor{AppBand}\multicolumn{4}{@{}l}{\textbf{B12 --- Winter Tai Chi sequence}}\\*
\rowcolor{AppStripe}
\texttt{image}\newline 027
& 2026-01-02 04:10:11\newline Friday; festival: none\newline Topic: Tai Chi\newline POI: none\newline 2 likes; 0 comments
& \textbf{Post:} ``Winter mornings in Chongqing are crisp! At 4 AM, the park is quiet and perfect for Tai Chi. Remember: keep your spine straight and breathe deeply to circulate qi. It's never too late to strengthen your balance and peace of mind.''\newline
\textbf{Payload:} 9 images; video count: none. Descriptions trace a complete dawn practice through mist, frost, stable foot placement, relaxed hands, and a quiet finishing pose.
& The multi-image sequence makes disciplined routine and age-appropriate persistence observable without relying on a single portrait. \\
\rowcolor{AppBand}\multicolumn{4}{@{}l}{\textbf{B13 --- Strength interpreted through music}}\\*
\texttt{share\_}\newline\texttt{music}\newline 029
& 2026-01-11 14:22:33\newline Sunday; festival: none\newline Topic: music\newline POI: none\newline 19 likes; 3 comments
& \textbf{Post:} ``The winter sun warms my back as I listen to `Hero.' Its melody carries the weight of years, reminding me that true strength lies in a gentle heart.''\newline
\textbf{Payload:} \textit{Hero} (translated title), artists Zheng Jun and Mo Hualun; the supplement associates the recording with a Gao Xiaosong works collection. Media: none.
& A shared song is interpreted through resilience and gentleness, reinforcing the persona's compassionate rather than performative notion of strength. \\
\rowcolor{AppBand}\multicolumn{4}{@{}l}{\textbf{B14 --- Two generations in the park}}\\*
\rowcolor{AppStripe}
\texttt{video}\newline 032
& 2026-01-19 09:29:52\newline Monday; festival: none\newline Topic: parenting\newline POI: none\newline 0 likes; 0 comments
& \textbf{Post:} ``Morning light on the park grass. Watching my grandson chase pigeons, I see the same joy his father once had. Time flows like the Yangtze, gentle yet unstoppable.''\newline
\textbf{Payload:} Image count: none; 1 video describing an elderly woman near a bench as a young man in his twenties runs across a frost-covered lawn.
& The adult grandson is interpreted through remembered parenthood, extending the profile's intergenerational continuity across time. \\
\rowcolor{AppBand}\multicolumn{4}{@{}l}{\textbf{B15 --- Shared television afternoon}}\\*
\texttt{video}\newline 044
& 2026-05-11 14:32:13\newline Monday; festival: none\newline Topic: Korean drama\newline POI: none\newline 4 likes; 1 comment
& \textbf{Post:} ``My grandson picked this Korean drama for us to watch. At 24, he still loves spending afternoons with his old grandma. So proud of his kind heart!''\newline
\textbf{Payload:} Image count: none; 1 video describing an elderly woman and a young man seated together watching a subtitled Korean drama.
& The shared medium broadens the history beyond traditional activities while retaining the warm, reciprocal grandparent--grandchild relationship. \\
\end{longtable}
\endgroup

\begingroup
\footnotesize
\renewcommand{\arraystretch}{1.08}
\setlength{\tabcolsep}{3.4pt}
\setlength{\LTleft}{0pt}
\setlength{\LTright}{0pt}
\begin{longtable}{@{}p{0.09\linewidth}p{0.20\linewidth}p{0.46\linewidth}p{0.19\linewidth}@{}}
\caption{Cross-modal post records for Persona C (Steady Hearth).}
\label{tab:post-showcase-steady-hearth}\\
\toprule
\textbf{Type / ID} & \textbf{Time, context, engagement} & \textbf{Authored text and payload} & \textbf{Persona realization} \\
\midrule
\endfirsthead
\endhead
\endfoot
\bottomrule
\endlastfoot
\rowcolor{AppBand}\multicolumn{4}{@{}l}{\textbf{C1 --- Work--family continuity}}\\*
\texttt{text}\newline 021
& 2025-12-16 14:12:57\newline Tuesday; festival: none\newline Topic: seasonal greetings\newline POI city: Weifang\newline POI name: media workplace (coarsened)\newline 31 likes; 10 comments
& \textbf{Post:} ``Just finished a quick lunch at the office. The winter sun is weak today, but seeing my granddaughter's drawing of snowflakes on my desk warms my heart. Can't wait to pick her up and make hot pear soup for everyone tonight. Winter in Mihe is cold, but our home is cozy.''\newline
\textbf{Payload:} Media/share: none.
& Office work, kindergarten pickup, family cooking, season, and a warm domestic register are integrated in one record. \\
\rowcolor{AppBand}\multicolumn{4}{@{}l}{\textbf{C2 --- Intergenerational overnight care}}\\*
\rowcolor{AppStripe}
\texttt{image}\newline 003
& 2025-06-06 03:14:46\newline Friday; festival: none\newline Topic: diary\newline POI: none\newline 22 likes; 3 comments
& \textbf{Post:} ``3:14 AM and the house is finally quiet. My granddaughter had a fever tonight, so her mom and I took turns watching her. She's sleeping soundly now. I'll brew some ginger tea for myself before heading back to bed. These late nights remind me how fast time flies; my son was once the one I stayed up with, and now I'm doing it for his little girl. Summer nights are warm, but the worry keeps you alert. Goodnight, world.''\newline
\textbf{Payload:} 1 image; video count: none. The description shows a dim living room, amber tea, a digital clock reading 03:14, and a summer blanket.
& Overnight care links past parenting to present grandparenting and makes multigenerational cooperation observable. \\
\rowcolor{AppBand}\multicolumn{4}{@{}l}{\textbf{C3 --- Professional reflection}}\\*
\texttt{video}\newline 006
& 2025-07-08 01:59:29\newline Tuesday; festival: none\newline Topic: work\newline POI: none\newline 0 likes; 0 comments
& \textbf{Post:} ``It's late, but the office is finally quiet. After decades in media, I've learned that the best decisions often come not in the rush of the day, but in these still moments. Watching the city sleep from my window, everything feels clear and in its place.''\newline
\textbf{Payload:} Image count: none; 1 video describing a dim office, desk lamp, papers, dark monitor, rain-streaked window, and city lights.
& ``Decades in media'' grounds the occupation, while deliberate judgment and quiet reflection align with steady, responsible leadership. \\
\rowcolor{AppBand}\multicolumn{4}{@{}l}{\textbf{C4 --- Quiet-night music share}}\\*
\rowcolor{AppStripe}
\texttt{share\_}\newline\texttt{music}\newline 019
& 2025-12-02 02:59:56\newline Tuesday; festival: none\newline Topic: music\newline POI city: Mihe\newline POI name: home (coarsened)\newline 4 likes; 0 comments
& \textbf{Post:} ``Can't sleep, so I'm enjoying this classic ballad. The quiet night makes the lyrics hit even harder.''\newline
\textbf{Payload:} \textit{Wo Xihuan (Ballad Version)} (transliterated title), artist Tao Zhe; the supplement identifies the track as a Tao Zhe recording from \textit{Ultrasound: The Road of Music 1997--2003}. Media: none.
& Nocturnal listening extends the profile's quiet-reflection style beyond family episodes while preserving topical variation. \\
\rowcolor{AppBand}\multicolumn{4}{@{}l}{\textbf{C5 --- Employment stability through a family lens}}\\*
\texttt{share\_}\newline\texttt{article}\newline 010
& 2025-09-21 09:40:07\newline Sunday; festival: none\newline Topic: reading\newline POI city: Mihe\newline POI name: withheld\newline 12 likes; 2 comments
& \textbf{Post:} ``Sunday morning and I'm reading about construction job hopping? My son's been in this industry for years, always stressed about cash flow. Just want him to find a stable spot so he can actually enjoy time with his daughter.''\newline
\textbf{Payload:} \textit{Which Employer Is More Reliable for Cost Engineers Changing Jobs?} (translated title); preview recommends evaluating prospective construction employers' cash-flow health. Artist/media: none.
& The article is interpreted through her adult son's stability and time with his daughter, linking financial prudence to family welfare. \\
\rowcolor{AppBand}\multicolumn{4}{@{}l}{\textbf{C6 --- Team-safety planning}}\\*
\texttt{video}\newline 005
& 2025-07-03 12:15:26\newline Thursday; festival: none\newline Topic: group outings\newline POI: none\newline 24 likes; 0 comments
& \textbf{Post:} ``Lunch break tip for my team: When organizing group outings in this July heat, always schedule activities before 10 AM or after 4 PM. Hydration stations are non-negotiable. Safety first, fun second!''\newline
\textbf{Payload:} Image count: none; 1 video describing heat above a sunny park path and a table prepared with water bottles and umbrellas.
& The team address, explicit scheduling rule, and safety priority expose a practical supervisory stance and responsible leadership. \\
\rowcolor{AppBand}\multicolumn{4}{@{}l}{\textbf{C7 --- Kindergarten creativity}}\\*
\rowcolor{AppStripe}
\texttt{text}\newline 007
& 2025-07-11 13:11:05\newline Friday; festival: none\newline Topic: seasonal greetings\newline POI city: Mihe\newline POI name: kindergarten (coarsened)\newline 3 likes; 1 comment
& \textbf{Post:} ``My little granddaughter drew this for me at kindergarten! So proud of her creativity.''\newline
\textbf{Payload:} Media/share: none.
& Kindergarten context and explicit grandparental pride instantiate the everyday family-anchor role. \\
\rowcolor{AppBand}\multicolumn{4}{@{}l}{\textbf{C8 --- National Day meal preparation}}\\*
\texttt{video}\newline 011
& 2025-10-02 18:44:50\newline Thursday; National Day\newline Topic: home cooking\newline POI: none\newline 34 likes; 5 comments
& \textbf{Post:} ``National Day feast prep! Trying to sneak a veggie into the pot before the meat lovers notice.''\newline
\textbf{Payload:} Image count: none; 1 video describing hands stirring vegetables and meat in a steaming wok on a home stove.
& Festival cooking realizes a declared interest and household responsibility, while mild humor sustains a warm family register. \\
\rowcolor{AppBand}\multicolumn{4}{@{}l}{\textbf{C9 --- Dumplings with the granddaughter}}\\*
\rowcolor{AppStripe}
\texttt{text}\newline 020
& 2025-12-05 17:49:27\newline Friday; festival: none\newline Topic: foodie life\newline POI: none\newline 8 likes; 0 comments
& \textbf{Post:} ``Just finished making dumplings for dinner. My 4-year-old granddaughter helped fold them! Does anyone have a good recipe for a warm winter soup to go with them? Need ideas!''\newline
\textbf{Payload:} Media/share: none.
& Shared meal preparation makes traditional cooking and age-consistent intergenerational participation simultaneously observable. \\
\rowcolor{AppBand}\multicolumn{4}{@{}l}{\textbf{C10 --- Playful kindergarten commute}}\\*
\texttt{image}\newline 022
& 2025-12-25 09:38:58\newline Thursday; Christmas\newline Topic: daily\newline POI: none\newline 0 likes; 1 comment
& \textbf{Post:} ``Merry Christmas! My 4yo granddaughter insisted I wear these reindeer antlers to drop her off at kindergarten. At 54, I'm too old for this, but her laugh is worth the embarrassment. Now heading to work looking like a festive deer!''\newline
\textbf{Payload:} 9 images; video count: none. Descriptions connect the antlers and laughing child with breakfast, the kindergarten route, public transit, and the author's office desk.
& Age, kindergarten drop-off, work transition, and affectionate humor integrate the profile's household and occupational routines. \\
\rowcolor{AppBand}\multicolumn{4}{@{}l}{\textbf{C11 --- Music across generations}}\\*
\texttt{text}\newline 024
& 2026-01-16 16:16:42\newline Friday; festival: none\newline Topic: guitar\newline POI: none\newline 35 likes; 11 comments
& \textbf{Post:} ``Just finished work and picked up my granddaughter. On the drive home, we heard a street musician playing guitar. She was so mesmerized! It reminded me of how music touches every generation. Now heading home to cook dinner for the family.''\newline
\textbf{Payload:} Media/share: none.
& Work, pickup, shared listening, and meal preparation are joined in one transition from professional to family responsibility. \\
\rowcolor{AppBand}\multicolumn{4}{@{}l}{\textbf{C12 --- Winter lantern evening}}\\*
\rowcolor{AppStripe}
\texttt{image}\newline 025
& 2026-02-12 18:43:29\newline Thursday; festival: none\newline Topic: family\newline POI: none\newline 19 likes; 3 comments
& \textbf{Post:} ``Lanterns glow soft against the winter dusk. Four years old, she points at the moon, unaware she is the brightest light in our sky.''\newline
\textbf{Payload:} 9 images; video count: none. Descriptions show lanterns, the four-year-old child, tangyuan, gloved intergenerational hands, a family group, and winter street and window details.
& The four-year-old child's age, seasonal domestic cues, and affectionate metaphor preserve the warm Family Anchor voice. \\
\rowcolor{AppBand}\multicolumn{4}{@{}l}{\textbf{C13 --- Traditional broth}}\\*
\texttt{text}\newline 034
& 2026-04-24 10:12:51\newline Friday; festival: none\newline Topic: cooking traditional meals\newline POI: none\newline 22 likes; 4 comments
& \textbf{Post:} ``Spring light filters through the kitchen window as I simmer the broth. The scent of ginger and star anise fills the air, a quiet poem for my family's dinner.''\newline
\textbf{Payload:} Media/share: none.
& Traditional cooking is expressed through an unhurried sensory description consistent with the profile's gentle evening-reflection style. \\
\rowcolor{AppBand}\multicolumn{4}{@{}l}{\textbf{C14 --- Late snack with her son}}\\*
\rowcolor{AppStripe}
\texttt{image}\newline 036
& 2026-05-12 20:26:05\newline Tuesday; festival: none\newline Topic: midnight diner\newline POI: none\newline 24 likes; 4 comments
& \textbf{Post:} ``Late night snack with my son after he picked up his daughter. Quiet moments like these warm the heart. Wishing everyone peace and good health.''\newline
\textbf{Payload:} 6 images; video count: none. Descriptions show noodle soup, tea, two place settings, and a nearby child's plate in a dim home kitchen.
& The pickup chain and two-place meal make cooperation between grandmother, adult son, and granddaughter visible without enumerating the household. \\
\rowcolor{AppBand}\multicolumn{4}{@{}l}{\textbf{C15 --- Child-assisted lunch preparation}}\\*
\texttt{video}\newline 037
& 2026-05-21 11:40:57\newline Thursday; festival: none\newline Topic: emotions\newline POI: none\newline 28 likes; 4 comments
& \textbf{Post:} ``My 4-year-old granddaughter tried to `help' me chop veggies for lunch. Let's just say the kitchen looks like a tomato exploded!''\newline
\textbf{Payload:} Image count: none; 1 video describing scattered tomato pieces, a child's plastic knife, and an adult reaching for a towel while laughing.
& The closing example combines cooking, safe participation by a four-year-old, and affectionate humor in an ordinary family scene. \\
\end{longtable}
\endgroup

\FloatBarrier

\begin{center}
\captionof{table}{Schema of a SocialEnv post record and its use by the five-tool retrieval pipeline. Shared fields occur across modalities; conditional fields supply visual, shared-entity, geographic, or engagement evidence to the corresponding operator.}
\label{tab:post-record-composition}
\footnotesize
\renewcommand{\arraystretch}{1.08}
\setlength{\tabcolsep}{4pt}
\begin{tabular}{@{}p{0.18\linewidth}p{0.15\linewidth}p{0.35\linewidth}p{0.25\linewidth}@{}}
\toprule
\textbf{Field group} & \textbf{Presence} & \textbf{Stored fields} & \textbf{Consuming operator and role} \\
\midrule
\rowcolor{AppBand}\multicolumn{4}{@{}l}{\textbf{Shared record structure}}\\
Identity and authorship
& All posts
& Source-local \texttt{post\_id}; directory-/loader-level author association.
& \texttt{select\_by\_person} resolves author constraints through the association; identifiers maintain linkage but are not semantic evidence. \\
\rowcolor{AppStripe}
Temporal metadata
& All posts
& Timestamp, weekday, and nullable/list-valued festival field.
& \texttt{select\_by\_time} evaluates date, weekday, festival, and time-of-day predicates. \\
Modality and authored text
& All posts; text may be empty
& One of five type enums and the user-authored caption or body.
& \texttt{select\_by\_content} applies the exact type gate before semantic matching over observable content. \\
\rowcolor{AppStripe}
Topic annotation
& All posts
& Generator-assigned topic list.
& Generation metadata; withheld as an independent semantic cue from \texttt{select\_by\_content}. \\
\rowcolor{AppBand}\multicolumn{4}{@{}l}{\textbf{Modality-specific semantic evidence}}\\
Visual payload
& Image/video only
& Image/video count and textual media-description array.
& \texttt{select\_by\_content} can match non-text evidence without requiring raw media in the local release. \\
\rowcolor{AppStripe}
Shared entity
& Music/article only
& Shared title, optional artist, and descriptive supplement or article preview.
& \texttt{select\_by\_content} receives the referenced entity and its observable supporting text. \\
\rowcolor{AppBand}\multicolumn{4}{@{}l}{\textbf{Optional context and terminal presentation}}\\
Geographic provenance
& Optional
& Post-level POI city and name, distinct from author residence.
& \texttt{select\_by\_location} evaluates explicit place constraints. \\
\rowcolor{AppStripe}
Engagement
& All posts
& Like count and comment count.
& \texttt{finalize} applies requested likes, comments, interactions, or importance ordering after relevance filtering. \\
\bottomrule
\end{tabular}
\end{center}

\end{document}